\documentclass[preprint,12pt]{elsarticle}

\usepackage{graphicx}
\usepackage{amssymb}
\usepackage{amsthm}
\usepackage{lineno}
\usepackage{amsmath}
\usepackage{listings}
\usepackage{textcomp}
\usepackage{xcolor}
\usepackage{color}
\usepackage{courier}
\usepackage{subfig}

\lstdefinestyle{customc}{
  belowcaptionskip=1\baselineskip,
  breaklines=true,
  frame=L,
  xleftmargin=\parindent,
  language=C,
  showstringspaces=false,
  basicstyle=\footnotesize\ttfamily,
  keywordstyle=\bfseries\color{green!40!black},
  commentstyle=\itshape\color{purple!40!black},
  identifierstyle=\color{blue},
  stringstyle=\color{orange},
}

\journal{Computer Physics Communications}

\begin{document}

\begin{frontmatter}

\title{Modular, general purpose ODE integration package to solve large number of independent ODE systems on GPUs}

\author[bme]{Ferenc Heged\H us\corref{cor1}}
\ead{fhegedus@hds.bme.hu}

\cortext[cor1]{Corresponding author}
\address[bme]{Budapest University of Technology and Economics, Faculty of Mechanical Engineering, Department of Hydrodynamic Systems,  P.O. Box 91, 1521 Budapest, Hungary}


\begin{abstract}
A general purpose, modular program package for the integration of large number of independent ordinary differential equation systems capable of using professional graphics cards is presented. The available numerical schemes are the explicit and adaptive Runge--Kutta--Cash--Karp algorithm and the explicit fourth order Runge--Kutta method with fixed time step. In order to harness the huge processing power of graphics cards, the intermediate points of the computed trajectories are not stored. As a compensate, with pre-declared device functions, the required special features or properties of a solution can be easily extracted and stored each into a dedicated variable. For instance, the maximum and minimum values and/or their time instances. Event handling is also incorporated into the package in order to detect special points which can be stored as well. Moreover, again with pre-declared device function calls at such special points, the efficient handling of non-smooth dynamics---e.g. impact dynamics---is possible. Several test cases are presented to demonstrate the flexibility of the pre-declared device functions and the strength of the program package. The applied models are the simple Duffing oscillator, the more complex Keller--Miksis equation known in bubble dynamics, and a system describing the behaviour of a pressure relief valve that can exhibit impact dynamics.
\end{abstract}


\end{frontmatter}


\section{Introduction}

In many physical problems, the governing equations describing the dynamics are initial value problems of first order ordinary differential equation (ODE) systems. Even partial differential equations can be decomposed into a large system of ODEs via suitable spatial discretization \cite{Boyd2001,Canuto2006}. The size of such systems (number of the involved equations) can vary between orders of magnitude. Among the simplest models (second or third order systems), one can find the classic Duffing \cite{Bonatto2008b,Englisch1991,Gilmore1995,Kao1987,Kozlowski1995,Parlitz1985,Wang1992}, Morse \cite{Knop1990,Scheffczyk1991}, Toda \cite{Kurz1988,Goswami1996,Goswami1998,Goswami2011} and Lorentz \cite{Goswami2007,Meucci2008,Goswami2008} equations which are extensively studied for many decades from non-linear dynamical point of view to establish bifurcation theories. Examples for medium-sized systems are the complex model of a pressure relief vale \cite{Hos2014,Hos2015} (order greater than $5$); single bubble dynamical model including partial differential equations or chemistry \cite{Yasui2008,Stricker2011,Yasui2012,Stricker2014,Hao1999,Hegedus2013b} (order greater than $20$); globally \cite{Wiesenfeld1989,Kaneko1990,Luther2000,Pikovsky2001,Yasui2010,Behnia2013b} or diffusionally \cite{Osipov1998,Fenton2005,Shabunin2009,Parlitz2011,Berg2011} coupled low order identical models in which a complete system can include several hundreds or even thousands of equations. Finally, extremely large system sizes can arise from the spatial discretization of a hydrodynamical problem \cite{Canuto2007,Niemeyer2014} or the simulation of chemical kinetics in reactive flows \cite{Stone2013,Sewerin2015,Imren2016,Curtis2017,Stone2018}, where the system size can be in the order of million.

Since the time evolution of a differential equation (initial value problem) is serial in nature, the application of massively parallel computation techniques is not trivial. One way is to use single instruction multiple data (SIMD) approach if the system is large enough and composed mostly by equations of the same form to distribute the tasks (e.g. right-hand side evaluations) within a single time step to several parallel threads \cite{Berg2011}. A special version of this method in GPU programming is often called \textit{per-block} approach \cite{Stone2018}. Another way is to perform parameter studies where the objective is to simulate large number of \textit{independent} and identical systems each having a different parameter set (and/or initial conditions) \cite{deSouza2012,Francke2013,Celestino2014,Gallas2015}. Throughout this paper, a recently developed ODE integration package is introduced following the second approach. The code is capable to exploit the high processing power of professional, general purpose graphics cards (GPUs). Due to their single instruction multiple thread (SIMT) architecture \cite{CUDAProgGuide2018}, the parameter sweeping problem is well-suited for GPUs, in which each independent ODE system is assigned to a single thread (\textit{per-thread} approach \cite{Stone2018}). Therefore, usually several thousands of systems have to be solved simultaneously in order to fully utilize a GPU. The source codes are written in C++ and CUDA C, and they are available as supplementary materials of this manuscript. During the implementation, special care was taken to be able to give the required user functions (e.g. the right hand side of the ODEs) as simple as possible and similar as in the case of MATLAB, for details see Sec.\,\ref{description_source_code}.

Before proceeding further, the scientific importance of huge parameter scans must be emphasized. If a system have many parameters then the number of the required simulations can blow up really fast. For example, a simple four dimensional parameter space with a moderate resolution of $100$ means altogether $100$ million ($100^4$) initial value problems. Similar computations have already been employed to investigate high resolution bi-parametric topologies of the bifurcation structure of non-linear systems \cite{Bonatto2008a,Freire2009,Gallas2010,Cabeza2013,Medeiros2011,Medrano2014,Rocha2015,Horstmann2017,Manchein2017,daSilva2017,Prants2017}; including studies of the present author \cite{Klapcsik2017,Hegedus2018}. Similarly, large parameter scans may be the only option in optimization problems of parameter fitting to measured data, where the corresponding error function have very non-smooth nature \cite{AlOmari2013,Kovac2018}. Although the above described ``brute force'' approach may seem to be a waste of resources in some cases, a high resolution, multi-dimensional pattern of a quantity can help identifying new features of non-linear systems which would be hidden otherwise. A good example is the recently published paper of Heged\H{u}s et al. \cite{Hegedus2018}, in which a new non-feedback technique to control multi-stability is introduced and described. The original purpose of the study was to investigate the collapse-like behaviour of a single spherical bubble related to the topic of acoustic cavitation and sonochemistry.

In order to accomplish the scan of millions of systems, the numerical code must be very fast and efficient. Therefore, the main concept is not to store every points of the trajectories, but only the endpoint and some special points of a simulation stopped either via event handling or by reaching a prescribed final time instance. In this way, the slow global memory operations and copy through PCI-E bus can be minimized. It is also mandatory from storage capacity viewpoint, since simulating large number of systems can easily fill-up an entire hard disk. In applications, many features of a particular solution might be required. For instance, the maximum value of a solution to create amplification diagrams, both the maximum and minimum radii in bubble dynamics to calculate the compression ratio, or special points detected by event handling in case of impact dynamics. In the introduced software package, flexible pre-declared device functions (e.g. for event handling) are responsible for the extraction of such information from the solutions. The proper usage of these functions are discussed in details throughout the present study. In this way, only the required special properties are transferred back to the host (main memory of the CPU) from the device (global memory of the GPU) after the end of a simulation instead of the whole trajectories. Although many other implementations (on GPUs) have been already published during the last years \cite{Shi2012,Murray2012,Dindar2013,Le2013,Demidov2013,Brock2015,Fazanaro2016}, \textit{according to the best knowledge of the author, such a general technique to handle large number of independent ODEs is still not available in the literature}.

In the following, the employed models during the tests are introduced in Sec.\,\ref{test_models}; the available numerical schemes, the concept of event handling and the storage of special features are discussed in general in Secs,\,\ref{integration_algorithms}, \ref{concept_event_handling} and \ref{concept_accessories}, respectively. The detailed description of the source code is given in Sec.\,\ref{description_source_code}, while the strength of the program is demonstrated through test cases in Sec.\,\ref{test_cases}. Finally, the paper is closed with a Conclusion (Sec.\,\ref{conclusion}).


\section{The test models} \label{test_models}

The flexibility and efficiency of the code are demonstrated on several test cases employing three different mathematical models. The first model (Duffing equation) is chosen due to its simplicity and its extensive usage in non-linear dynamics. The second model describes the dynamics of the radial oscillation of a single spherical gas bubble placed in liquid water driven by dual-frequency excitation. The equation itself is only a second order ODE; however, its from is very complex and contain many parameters. The last system describes the dynamics of a pressure relief valve, in which impact dynamics is possible between the valve body and the seat.

\subsection{The Duffing equation}

The duffing equation is a double well potential non-linear oscillator describing a periodically forced steel beam deflected between two magnets \cite{Duffing1918}. It is a second-order ordinary differential equation written as
\begin{align}
\dot{y}_1  &= y_2, \label{duffing_equation_1} \\
\dot{y}_2  &= \delta y_1 - y^3 - k y_2 + B \cos(\omega t) = F_2, \label{duffing_equation_2}
\end{align}
where $k$ is the damping factor and $B=0.3$ is the amplitude of the harmonic excitation. For the sake of simplicity, the stiffness of the beam $\delta=1$ and the angular frequency of the excitation $\omega=1$ are unity.

In Sec.\,\ref{test_duffing}, Eqs.\,\eqref{duffing_equation_1}-\eqref{duffing_equation_2} are used to calculate simple bifurcation diagrams by defining the global Poincar\'e section as the integer multiple of the period of the excitation; that is, the trajectories are sampled at time instances $t_n = n \cdot 2 \pi$ ($n=0,1,2,\ldots$). The control parameter is the damping coefficient $k$ with fixed excitation amplitude $B$. The simulations are repeated with the detection of the maximum and minimum values of $y_1$ between every points of the Poincar\'e section using two different techniques. In this way, for example, amplification diagrams can be constructed.

Lyapunov-exponent diagrams are also created. To do this in an efficient way, the linearized equations are attached to Eqs.\,\eqref{duffing_equation_1}-\eqref{duffing_equation_2} in polar coordinates following the concept of Parlitz and Lauterborn \cite{Parlitz1986}:
\begin{align}
\dot{y}_3  &= y_3( (1+g_1) \sin(y_4) \cos(y_4) + g_2 \sin(y_4)^2 ), \label{duffing_equation_3} \\
\dot{y}_4  &= -\sin(y_4)^2 + ( g_1 \cos(y_4) + g_2 \sin(y_4) ) \cos(y_4), \label{duffing_equation_4}
\end{align}
where $y_3$ and $y_4$ are the radius and angle of the linearized system\,\eqref{duffing_equation_3}-\eqref{duffing_equation_4}, respectively. The functions $g_1$ and $g_2$ are defined as
\begin{align}
g_1  &= \frac{\partial F_2}{\partial y_1}, \label{duffing_equation_g1} \\
g_2  &= \frac{\partial F_2}{\partial y_2}. \label{duffing_equation_g2}
\end{align}
Observe that system\,\eqref{duffing_equation_1}-\eqref{duffing_equation_2} is independent from system\,\eqref{duffing_equation_3}-\eqref{duffing_equation_4}; that is, there is a one-way coupling. The largest Lyapunov exponent can be easily obtained by averaging the linearized radius $y_3$ extracted at time instances $t_n$ of the Poincar\'e section:
\begin{equation}
\lambda_{\mathrm{max}} = \lim_{N\to\infty} \frac{1}{N} \sum_{n=1}^{N} y_3(t_n). \label{duffing_equation_lyap}
\end{equation}
It is important to reset $y_3$ to unity at every $t_n$, and to simulate (and discard) a suitably long transient phase before the calculation of the Lyapunov exponent to ensure the convergence of a particular trajectory to a stable state (attractor).

\subsection{The Keller--Miksis equation used in bubble dynamics and acoustic cavitation} \label{model_keller_miksis}

The second test model studied is the Keller--Miksis equation \cite{Lauterborn2010} describing the evolution of the radius of a gas bubble placed in a liquid domain and subjected to external excitation. The second-order ordinary differential equation reads
\begin{equation}\label{keller_miksis_1}
\left( 1-\frac{\dot{R}}{c_L} \right) R\ddot{R} + \left( 1-\frac{\dot{R}}{3c_L} \right) \frac{3}{2} \dot{R}^2 = \left( 1+\frac{\dot{R}}{c_L} + \frac{R}{c_L}\frac{d}{dt} \right) \frac{\left( p_L - p_{\infty}(t) \right)}{\rho_L},
\end{equation}
where $R(t)$ is the time dependent bubble radius; $c_L=1497.3\,\mathrm{m/s}$ and $\rho_L=997.1\,\mathrm{kg/m^3}$ are the sound speed and density of the liquid domain, respectively. The pressure far away from the bubble $p_{\infty}(t)$ is composed by static and periodic components
\begin{equation}\label{keller_miksis_2}
p_{\infty}(t) = P_{\infty} + P_{A1} \sin(\omega_1 t) + P_{A2} \sin(\omega_2 t + \theta),
\end{equation}
where $P_{\infty}=1\,\mathrm{bar}$ is the ambient pressure; and the periodic components have pressure amplitudes $P_{A1}$ and $P_{A2}$, angular frequencies $\omega_1$ and $\omega_2$, and a phase shift $\theta$. Such a dual-frequency driven gas bubble has paramount importance in the field of acoustic cavitation and sonochemistry \cite{Yasuda2007,Khanna2013,Brotchie2008,Zhang2015b,Zhang2017}.

The connection between the pressures inside and outside the bubble at its interface can be written as
\begin{equation}\label{keller_miksis_3}
p_G + p_V = p_L + \frac{2 \sigma}{R} + 4 \mu_L \frac{\dot{R}}{R},
\end{equation}
where the total pressure inside the bubble is the sum of the partial pressures of the non-condensable gas, $p_G$, and the vapour, $p_V=3166.8\,\mathrm{Pa}$. The surface tension is $\sigma=0.072\,\mathrm{N/m}$ and the liquid kinematic viscosity is $\mu_L=8.902^{-4}\,\mathrm{Pa\,s}$. The gas inside the bubble obeys a simple polytropic relationship
\begin{equation}\label{keller_miksis_4}
p_G = \left( P_{\infty} - p_V + \frac{2 \sigma}{R_E} \right) \left(\frac{R_E}{R}\right)^{3 \gamma},
\end{equation}
where the polytropic exponent $\gamma=1.4$ (adiabatic behaviour) and the equilibrium bubble radius is $R_E$.

The detailed description and the physical interpretation of Eqs.\,\eqref{keller_miksis_1}-\eqref{keller_miksis_4} is available in the previous papers of the authors \cite{Hegedus2013b,Hegedus2015}. It must be emphasized that the physical parameters of the system are the excitation properties: $P_{A1}$, $P_{A2}$, $\omega_1$, $\omega_2$, $\theta$ and the bubble size: $R_E$ (if the material properties and the static pressure are fixed). This large parameter space is reduced by setting the bubble size to $R_E=10\,\mathrm{\mu m}$ and the phase shift to $\theta=0$. In Sec.\,\ref{test_keller_miksis}, the achievable maximum expansion ratio of the bubble radius $(R_{\mathrm{max}}-R_E)/R_E$ (important measure of the efficiency of sonochemistry) of the solutions is investigated as high-resolution bi-parametric plots with excitation frequencies $\omega_1$ and $\omega_2$ as control parameters at fixed amplitudes $P_{A1}$ and $P_{A2}$. Observe that in this case the external forcing can be quasiperiodic; thus, special care have to be taken to handle the time domain during the simulations.

According to \cite{Hegedus2018}, system\,\eqref{keller_miksis_1}-\eqref{keller_miksis_4} is rewritten into a dimensionless form defined as
\begin{align}\label{dimensionless_system1}
\dot{y}_1  &= y_2,\\\label{dimensionless_system2}
\dot{y}_2  &= \frac{N_{\mathrm{KM}}}{D_{\mathrm{KM}}},
\end{align}
where the numerator, $N_{\mathrm{KM}}$, and the denominator, $D_{\mathrm{KM}}$, are
\begin{multline}\label{numerator}
N_{\mathrm{KM}} = \left( C_0 + C_1 y_2 \right) \left( \frac{1}{y_1} \right)^{C_{10}} - C_2 \left( 1 + C_9 y_2 \right) - C_3 \frac{1}{y_1} - C_4 \frac{y_2}{y_1} - \left( 1 - C_9 \frac{y_2}{3} \right) \frac{3}{2} y_2^2 \\
-\left( C_5 \sin(2 \pi \tau) + C_6 \sin(2 \pi C_{11} \tau + C_{12}) \right) \left( 1 + C_9 y_2 \right) \\
-y_1 \left( C_7 \cos(2 \pi \tau) + C_8 \cos(2 \pi C_{11} \tau + C_{12}) \right),
\end{multline}
and
\begin{equation}\label{denominator}
D_{\mathrm{KM}} = y_1 - C_9 y_1 y_2 + C_4 C_9.
\end{equation}

The coefficients are summarised as follows:
\begin{align}\label{coefficients}
C_0 &= \frac{1}{\rho_L} \left( P_{\infty} - p_V + \frac{2 \sigma}{R_E} \right) \left( \frac{2 \pi}{R_E \omega_1} \right)^2,\\
C_1 &= \frac{1-3\gamma}{\rho_L c_L} \left( P_{\infty} - p_V + \frac{2 \sigma}{R_E} \right) \frac{2 \pi}{R_E \omega_1},\\
C_2 &= \frac{P_{\infty} - p_V}{\rho_L} \left( \frac{2 \pi}{R_E \omega_1} \right)^2,\\
C_3 &= \frac{2 \sigma}{\rho_L R_E} \left( \frac{2 \pi}{R_E \omega_1} \right)^2,\\
C_4 &= \frac{4 \mu_L}{\rho_L R_E^2} \frac{2 \pi}{\omega_1},\\
C_5 &= \frac{P_{A1}}{\rho_L} \left( \frac{2 \pi}{R_E \omega_1} \right)^2,\\
C_6 &= \frac{P_{A2}}{\rho_L} \left( \frac{2 \pi}{R_E \omega_1} \right)^2,\\
C_7 &= R_E \frac{\omega_1 P_{A1}}{\rho_L c_L} \left( \frac{2 \pi}{R_E \omega_1} \right)^2,\\
C_8 &= R_E \frac{\omega_1 P_{A2}}{\rho_L c_L} \left( \frac{2 \pi}{R_E \omega_1} \right)^2,\\
C_9 &= \frac{R_E \omega_1}{2 \pi c_L},\\
C_{10} &= 3\gamma,\\
C_{11} &= \frac{\omega_2}{\omega_1},\\
C_{12} &= \theta.
\end{align}
Observe that from the implementation point of view, the number of the parameters of the system is $13$ ($C_0$ to $C_{12}$). Therefore, the aforementioned physical parameters and the appearing systems coefficients as parameters must be clearly separated in the code. Although the usage of the coefficients $C_{0-12}$---instead of the physical parameters---requires additional storage capacity and global memory load operations, it can significantly reduce the necessary computations (if the coefficients are precomputed).

\subsection{A simple model to described the impact dynamics of a pressure relief valve} \label{model_pressure_relief_valve}

The last test case describes the behaviour of a pressure relief valve which can exhibit impact dynamics. According to \cite{Hos2012}, the dimensionless governing equations are
\begin{align}
\dot{y}_1  &= y_2, \label{pressure_relief_valve_1} \\
\dot{y}_2  &= -\kappa y_2 - (y_1+\delta) + y_3, \label{pressure_relief_valve_2} \\
\dot{y}_3  &= \beta ( q - y_1 \sqrt{y_3} ), \label{pressure_relief_valve_3}
\end{align}
where $y_1$ and $y_2$ are the displacement and velocity of the valve body, respectively. $y_3$ is the pressure inside the reservoir chamber to where the pressure relief valve is connected. The fixed parameters in the system are as follows: $\kappa=1.25$ is the damping coefficient, $\delta=10$ is the precompression parameter and $\beta=20$ is the compressibility parameter. The control parameter during the simulations is the dimensionless flow rate $q$, for details see Sec.\,\ref{test_impact_dynamics}.

In Eqs\,\eqref{pressure_relief_valve_1}-\eqref{pressure_relief_valve_3}, the zero value of the displacement ($y_1=0$) means that the valve body is in contact with the seat of the valve. If the velocity of the valve body $y_2$ has a non-zero, negative value at this point, the following impact law
\begin{align}
y_1^+  &= y_1^- = 0, \label{impact_law_1} \\
y_2^+  &= -r y_2^-, \label{impact_law_2} \\
y_3^+  &= y_3^- \label{impact_law_3}
\end{align}
is applied. The Newtonian coefficient of restitution $r=0.8$ approximates the loss of energy of the impact. In Sec.\,\ref{test_impact_dynamics}, it shall be shown that by applying multiple event handling together with a special ``action function'' call at the impact detection, system\,\eqref{pressure_relief_valve_1}-\eqref{pressure_relief_valve_3} can be handled very efficiently.


\section{The integration algorithms} \label{integration_algorithms}

In the software package, two explicit integration algorithms can be selected. The first one is the adaptive Runge--Kutta--Cash--Karp method \cite{Hairer1993} that uses a fifth and a forth order accurate solutions to calculate the local error and estimate the step size. This algorithm is feasible and fast in most of the cases. The second option is the fourth order Runge--Kutta solver \cite{Hairer1993} with fixed time step size. If the time scales of a solution does not vary significantly, this algorithm can be faster due to the omitted logic of error handling.

It must be emphasized that the aforementioned algorithms are known to be suitable only for non-stiff problems. Otherwise, implicit schemes are usually mandatory. Employing GPUs, however, this distinction is not self-evident. Due to the single instruction multiple thread (SIMT) architecture, each thread (a single ODE system) in a warp (basic thread organization unit) must perform the same instruction but on different data. Therefore, complicated control flow logic can destroy the performance if the threads in a warp have different paths. Since in this case, the instructions can be done only in a serial fashion. This phenomenon is called thread divergence. This is why explicit algorithms produces much higher speed-up over CPUs for non-stiff problems (even by a factor of $100$ \cite{Niemeyer2014,Fazanaro2016}) compared to the implicit ones for stiff problems (usually less than a factor $10$ \cite{Stone2013}). This is a consequence of the much more simple control logic of explicit solvers causing much less hazard for thread divergence. Even if an implicit scheme can march with much larger time step, the GPU can performs orders of magnitude larger number of steps within the same time using an explicit technique \cite{Stone2018}. This justifies the usage of only non-stiff solvers in this first version of the software package since even for moderately stiff problems, explicit algorithms can outperform the implicit ones. For more details on the implementation and performance analysis of stiff to moderately stiff solvers, the reader is referred to the publications \cite{Murray2012,Dindar2013,Demidov2013,AlOmari2013,Stone2013,Niemeyer2014,Stone2014,Sewerin2015,Brock2015,Rodriguez2015,Fazanaro2016,Imren2016,Curtis2017,Stone2018}.


\section{The concept of event handling} \label{concept_event_handling}

In this section, the overview of the process of event handling is introduced. It is mandatory to define Poncar\'e section for autonomous systems, to define switching conditions for non-smooth dynamics, and it is important for the extraction of special properties of the solutions (keep in mind that intermediate points are not stored).

\begin{figure}[ht!]   
	\centering
		\includegraphics[width=8.6cm]{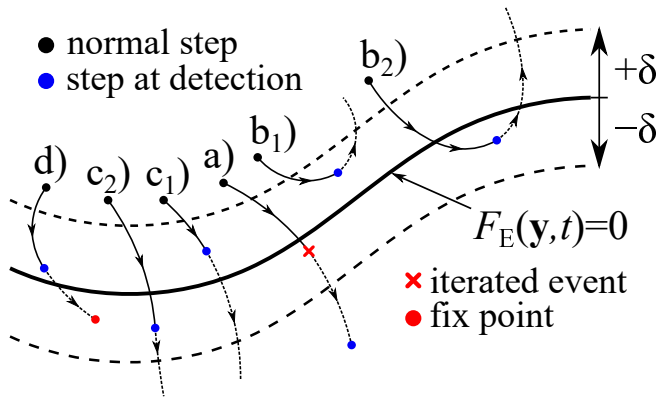}
	\caption{The concept of event handling and the possible scenarios of the event detection procedures.}
	\label{ConceptOfEventHandling}
\end{figure}

The concept is to set any number of user defined event functions describing subspaces in the state space of the system given in implicit form as
\begin{align}
F_{E1} (\textbf{y}, t) &= 0, \label{UDEF1} \\
F_{E2} (\textbf{y}, t) &= 0, \label{UDEF2} \\
\vdots \nonumber \\
F_{En} (\textbf{y}, t) &= 0, \label{UDEFn}
\end{align}
where $\textbf{y}$ is the state vector composed by the state variables. An example for a one dimensional event curve (solid black) in a two dimensional state space is shown in Fig.\,\ref{ConceptOfEventHandling}. Due to the finite precision of floating point numbers, a tolerance must be associated to each event function that defines a stripe $\pm \delta$ around the event curve called event zone here, see the dashed curves in Fig.\,\ref{ConceptOfEventHandling}. Since the tolerance is given in terms of the value of the event function, the thickness of the event zone is not necessarily fixed. The aim during a computation (in terms of event handling) is to monitor the crossings of the trajectory with the event zone, and to find at least one point within this zone.

In order to properly handle events, three different states of a particular trajectory must be distinguished during its life time. In case of \textit{normal state}, the trajectory is outside the range of any event zone. Check for event detection takes place only during the normal state. Figure\,\ref{ConceptOfEventHandling} summarizes the six possible configurations of an event detection. The black dots represent an accepted step still in the normal state right before an event detection while the blue dots show the next accepted step when a trajectory hits or crosses an event zone; here its state changes to \textit{detected state}.

The five event detection configurations $a$ to $c_2$ are shown in Fig.\,\ref{ConceptOfEventHandling}. Configuration $a$ has the highest probability if the prescribed tolerance is small and thus the trajectory steps over the whole event zone. To locate a point inside the zone marked by the red cross, a simple secant method is employed. In all the other configurations, if the event is detected, its corresponding point in the event zone is immediately found. Consequently, no additional iterations are necessary for the precise event location. After completing the event detection process (finding a point inside the event zone), the state of the trajectory is set to \textit{leaving state}. During this state, the trajectory must leave the event zone before switch back to normal state. It is mandatory to avoid multiple detection of the same event for the subsequent time steps still residing in the event zone. Observe that in cases $b_1$, $b_2$, $c_1$ and $c_2$, the state of the trajectory is immediately set to leaving state from normal state.

The last configuration marked by $d$ in Fig.\,\ref{ConceptOfEventHandling} represents a case where an equilibrium solution sits inside the event zone (red dot). If the trajectory converges to such a fix point, it will never leave the event zone. Therefore, a maximum number of time step can be specified to stop the simulation.

It must be emphasized again that complex control logic can causes divergent threads in a warp degrading the processing power of GPUs. Consequently, the number of the event functions should be as small as possible. If the determination of a certain feature of a solution with high precision is not crucial then a faster option is to use accessories functions discussed in more details in the next section. Finally, precise event location is possible only for a single event at a time. This means that if multiple events are detected at the same time (because of very close event curves and/or overlapping event zones), only the detected event having the largest serial number $n$ (specified last) will be located precisely. Thus, the order in which the functions are given via Eqs.\,\eqref{UDEF1}-\eqref{UDEFn} can be very important. The details to set-up event handling via pre-declared device functions are given in Sec.\,\ref{description_source_code}.


\section{The concept of accessories} \label{concept_accessories}

\begin{figure}[ht!]   
	\centering
		\includegraphics[width=8.6cm]{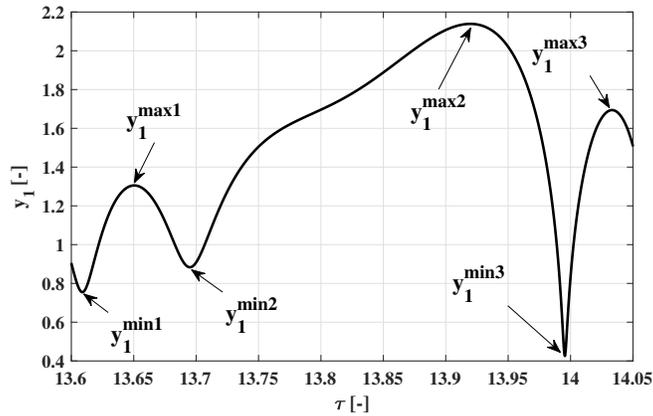}
	\caption{The concept of accessories introduced via local and global minima and maxima of a single component of a trajectory.}
	\label{ConceptOfAccessories}
\end{figure}

Accessories are the dedicated variables to store special properties of a trajectory, such as the local and/or the global maxima of a specific component of a solution. Figure\,\ref{ConceptOfAccessories} shows an example of the first component $y_1$ of a solution as a function of the dimensionless time $\tau$ (the underlying physics and the model is not important here). It can be clearly seen that the time-curve contains three local maxima and three local minima. In order to determine the global maxima, for instance, it is necessary to allocate only a single variable. Then during the integration of the system between the time instances $\tau=13.6$ and $\tau=14.05$, this specific variable can be continuously updated according to the actual value of an accepted time step. The updating process takes place via a special pre-declared ``ordinary'' accessory function called during the simulation at every successful step (the next section discuss its details). Similar procedure can be set-up to store the global minimum as well.

It must be emphasized that the control logic inside the ``ordinary'' accessory function body in the program is absolutely under the control of the user, in which the actual values of the state variables, the parameters and the actual time can be accessed. Therefore, assembling more complex properties is also possible depending on the requirements and the user imagination. For example, the time instant corresponding to the global maximum can be easily stored into another accessory variable updated together with the accessory of the global maximum as the actual time is known inside the function (see also Sec.\,\ref{details_accessories}).

Observe that the function presented in Fig.\,\ref{ConceptOfAccessories} contains several local maxima and minima. These special points can be stored only with the cooperation of event handling. First, a suitable event function (the derivative of $y_1$ is zero) must be defined to detect these extreme values. At every successful event detection, another pre-declared device function is called. It works similarly as the ``ordinary'' accessory function call with the extension of an event counter variable; that is, the local maxima or minima can be stored according to their serial number. For details of the implementation, the reader is referred again to Sec.\,\ref{details_accessories}; and for different examples through test cases to Sec.\,\ref{test_cases}.

As a final remark, the required number of accessories variables $N_A$ are allocated for each independent system. Therefore, if the number of the independent systems solved simultaneously is $N_T$ (subscript $T$ is referred to the number of threads running concurrently, see the next section for the details), the total number of the allocated accessories variables are $N_A \times N_T$ during a single run. In addition, the proper initialization of the values of the accessories is mandatory. For this purpose, a pre-declared initialization function is call only once at the beginning of each integration phase (introduced in Sec.\,\ref{details_initialization_finalization}). In case of the global maximum or minimum, for instance, the proper initialization is to use the initial conditions.


\section{Description of the source code} \label{description_source_code}

Throughout this section, the details of the implementation of the program package is discussed including the definition and the set-up of the whole problem, how to organize it into smaller chunks of tasks, and how to define properly all the necessary pre-declared device functions including the right-hand side of the ODE system. Since function pointers cannot be passed to a GPU kernel function, the pre-declaration of a handful of global device functions is inevitable to manage the integration process (some of them are already mentioned above). In order to avoid name conflicts in larger programs, these functions have unique and long names. Although these functions are pre-declared (the input arguments cannot be modified), the definition of the function body is absolutely under the control of the user. If a specific function is not necessary then its function body can be left empty. In this case, a compiler shall optimize out the whole function call from the code.

\subsection{Set-up the problem pool} \label{problem_pool}

\begin{figure}[ht!]   
	\centering
		\includegraphics[width=8.6cm]{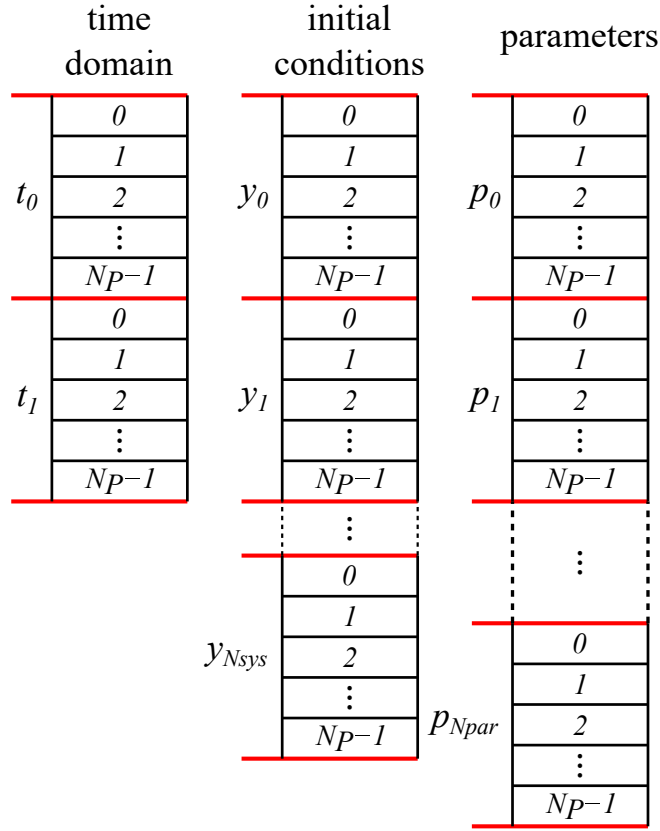}
	\caption{The proper structure and pattern of the problem pool.}
	\label{ProblemPool}
\end{figure}

The first objective is to properly create a pool of problem consisting of many independent systems of ODE. The general form of one system is
\begin{equation}\label{general_form_of_a_system}
\dot{\textbf{y}} = f \left( \textbf{y}, t; \textbf{p} \right), 
\end{equation}
where $\textbf{y}$ is the vector of the state variables, $\textbf{p}$ is the vector of the parameters and the integration time domain is $t \in (t_0, t_1)$. In the following, the number of the equations in a single system (system size) and the number of its parameters are denoted by $N_{sys}$ and $N_{par}$, respectively. During the simulations, one system is assigned to a single GPU thread. For a proper definition of an initial value problem, the time domain, the initial conditions of the state variables and the values of the parameters have to be specified for each system. If the number of the independent systems in the pool (problem size) is $N_P$ then the size of the required linear memory allocations are $2 \times N_P$, $N_{sys} \times N_P$ and $N_{par} \times N_P$ for the time domain, initial conditions and parameters, respectively (naturally, multiplied each by the size of type double):
\lstset{escapechar=@,style=customc}
\begin{lstlisting}
double* TimeDomain_Pool=AllocateHostMemory<double> \
	(2*ProblemSize);

double* InitialCondition_Pool=AllocateHostMemory<double> \
	(SystemDimension*ProblemSize);

double* Parameters_Pool=AllocateHostMemory<double> \
	(NumberOfParameters*ProblemSize);

double* Accessories_Pool=AllocateHostMemory<double> \
	(NumberOfAccessories*ProblemSize);
\end{lstlisting}
The function {\fontfamily{pcr}\selectfont \textcolor{blue}{AllocateHostMemory<>()}} do the allocation itself in the host memory via the {\fontfamily{pcr}\selectfont \textbf{\textcolor{blue}{new}}} command providing an ordinary pointer as a return value containing the address of the first element in the array; the code is not listed here. In the above code snippet, memory allocation for the accessories are also done. It is included only for completeness and it is necessary to concentrate only on the first three arrays in this section. Accessories are discussed in details during the subsequent sections. In order to achieve coalesced global memory operations/access on the device memory of the GPU(s), special care must be taken already during the filling of these arrays in the host memory. The proper data pattern is depicted in Fig.\,\ref{ProblemPool}. Due to the single instruction multiple thread (SIMT) approach of GPUs, $32$ threads (a warp) must perform the same instruction but on different data. This rule holds for the memory load/write operations as well; that is, if a variable is required in an instruction (e.g. $y_1$), then load operation is issued for $32$ number of $y_1$ values corresponding to the $32$ number of threads. To accelerate data transfer to/from the global memory, the memory is accessed via $32$, $64$ or $128$ byte transactions \cite{CUDAProgGuide2018}. If the required data are coalesced, the number of the required transactions can be minimized and the throughput can be maximized. In the worst case, $32$ number of $32$ byte memory transactions are required if the data is scattered in the global memory. Since global memory latency is very high ($600-800$) cycles, coalesced memory access pattern is mandatory to fully utilize the processing power of a GPU. Therefore, the $N_P$ number of values of the aforementioned variable $y_1$ must be stored consecutively in a memory block. All the other components of the state variable, parameters and time domains must be stored similarly. This explains the pattern shown in Fig.\,\ref{ProblemPool}, where the numbers inside the rectangles actually show the serial numbers of the independent ODE systems in the pool. It is the responsibility of the user to fill these arrays with valid data with a proper pattern.

Observe that every system has its own time coordinate with its own time domain ($t_0$ to $t_1$). This means that each system is integrated independently to each other. Naturally, this can cause thread divergence within a warp in case of the adaptive solver if the different systems require very different number of time step to complete the integration. In our experience, however, this approach is still better than using a common time coordinate and a common time step value for each system in a warp or a block. Possibly this is the technique suggested by Curtis et al. \cite{Curtis2017} to synchronize the time step adoption and avoid thread divergence. If the solutions contain very different time scales, and all the $32$ systems slow down at different time instances, then the whole simulation have to slow down unnecessarily many times. Parenthetically, the studies in the literature using the later technique (common time coordinate) Demidov et al. \cite{Demidov2013} reported that parameter studies on GPUs become faster only if the number of systems solved simultaneously is in the order of $10^5-10^6$. We shall see in Sec.\,\ref{test_cases} that with our method, a GPU can be highly utilized even with few thousands residing threads (systems).

\subsection{Initialize the solver object} \label{solver_object_initialization}

The problem size $N_P$ for a single parameter scan can be in the order of millions. However, sometimes it is not efficient or advisable to solve all the problems defined in a problem pool on a single GPU or as a single run. Therefore, the program package offers a special {\fontfamily{pcr}\selectfont \textbf{\textcolor{green!40!black}{class}}} and {\fontfamily{pcr}\selectfont \textbf{\textcolor{green!40!black}{struct}}} suitable for splitting the whole problem into smaller chunks of tasks and distribute them to different GPUs or solve them one after another on a single GPU. First, the system dimension $N_{sys}$, the number of the used threads in a grid $N_T$ (number of the independent systems solved simultaneously), the number of the system parameters $N_{par}$, the number of the used event functions $N_E$ and the number of the required accessories variables for a thread $N_A$ have to be specified via the {\fontfamily{pcr}\selectfont \textcolor{blue}{ParametricODESolverConfiguration}} structure as follows (the employed values are only examples):
\lstset{escapechar=@,style=customc}
\begin{lstlisting}
ParametricODESolverConfiguration ConfigurationSystem;
	ConfigurationSystem.SystemDimension     = 2;
	ConfigurationSystem.NumberOfThreads     = 7680;
	ConfigurationSystem.NumberOfParameters  = 21;
	ConfigurationSystem.NumberOfEvents      = 2;
	ConfigurationSystem.NumberOfAccessories = 2;
\end{lstlisting}
Next, a solver object (called {\fontfamily{pcr}\selectfont \textcolor{blue}{ScanSystem}} here) can be created with a simple declaration:
\begin{lstlisting}
ParametricODESolver ScanSystem(ConfigurationSystem);
\end{lstlisting}
The constructor of the {\fontfamily{pcr}\selectfont \textbf{\textcolor{green!40!black}{class}}} {\fontfamily{pcr}\selectfont \textcolor{blue}{ParametricODESolver}} performs all the necessary memory allocations both on the host (main memory) and the device (GPU memory). Moreover, the constant values of the Butcher tableau of the Runge--Kutta schemes are immediately copied into the constant memory of the selected GPU as well. The main advantage of the constant memory is that a single variable is broadcast to each thread in a single load operation. Keep in mind that due to the immediate memory allocations, the GPU device must be selected in advance via the {\fontfamily{pcr}\selectfont \textcolor{blue}{cudaSetDevice()}} function, before the object is created.

Utilizing a single GPU device, the creation of a single object is sufficient. In case of multiple GPUs, as many object have to be declared as the number of the devices. In this case, the proper order of the device selection and the object creation is extremely important to keep the data consistent in the global memory of each devices.

\subsection{Fill the solver object with a chunk of problems} \label{fill_the_solver_object}

\begin{figure}[ht!]   
	\centering
		\includegraphics[width=5.6cm]{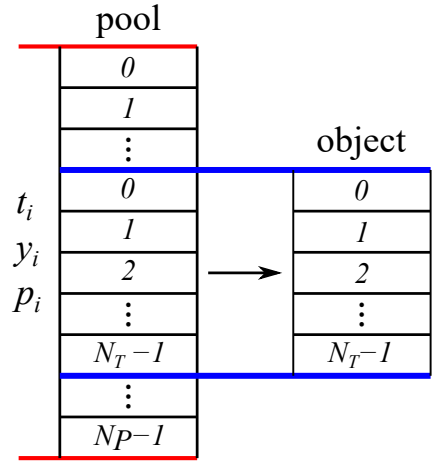}
	\caption{Copy a chunk of data of $N_T$ number of consecutive systems from the pool into the solver object with the {\fontfamily{pcr}\selectfont \textcolor{blue}{LinearSet()}} member function.}
	\label{CopyDataToObject}
\end{figure}

Although the creation of a {\fontfamily{pcr}\selectfont \textcolor{blue}{ParametricODESolver}} object performs all the necessary memory allocations, its internal storages have to be filled-up with valid data from the problem pool. There are two member functions responsible for this procedure. The first one copies the data of a consecutive set of systems from the pool into the object. This function is called {\fontfamily{pcr}\selectfont \textcolor{blue}{LinearSet()}}, which needs an input argument of a pre-defined type {\fontfamily{pcr}\selectfont \textcolor{blue}{LinearSetConfiguration}} structure:
\lstset{escapechar=@,style=customc}
\begin{lstlisting}
struct LinearSetConfiguration
{
	int CopyStartIndexInObject;
	int CopyStartIndexInPool;
	int PoolSize;
	int NumberOfElements;
	
	double* PoolTimeDomain;
	double* PoolActualState;
	double* PoolParameter;
	double* PoolAccessories;
	
	CopyModePossibilities CopyMode;
};
\end{lstlisting}
In this structure, the starting index (serial number of the systems) in the object (between $0$ and $N_T$) and in the pool (between $0$ and $N_P$) need to be specified first. Then the total number of the independent systems in the pool and the number of the systems (elements) whose properties are to be copied have to be set. Naturally, the pointers of the time domain, the initial conditions (actual state), the parameters and the accessories in the problem pool is necessary as well, see also Sec.\,\ref{problem_pool}. The {\fontfamily{pcr}\selectfont \textcolor{blue}{CopyModePossibilities}} field has five possible values: {\fontfamily{pcr}\selectfont \textbf{\textcolor{blue}{TimeDomain}}} to copy only the values of the time domain, {\fontfamily{pcr}\selectfont \textbf{\textcolor{blue}{ActualState}}} to copy only the initial conditions, {\fontfamily{pcr}\selectfont \textbf{\textcolor{blue}{Parameter}}} to copy only the parameters, {\fontfamily{pcr}\selectfont \textbf{\textcolor{blue}{Accessories}}} to copy only the accessories and finally {\fontfamily{pcr}\selectfont \textbf{\textcolor{blue}{All}}} to copy the values of all the aforementioned four properties. An example can be seen in the following code snippet:
\lstset{escapechar=@,style=customc}
\begin{lstlisting}
LinearSetConfiguration LinearCopySetup;

LinearCopySetup.CopyStartIndexInObject = 0;
LinearCopySetup.CopyStartIndexInPool   = 0;
LinearCopySetup.PoolSize         = ProblemSize;
LinearCopySetup.NumberOfElements = NumberOfThreads;
LinearCopySetup.PoolTimeDomain   = TimeDomain_Pool;
LinearCopySetup.PoolActualState  = InitialCondition_Pool;
LinearCopySetup.PoolParameter    = Parameters_Pool;
ActualCopySetup.PoolAccessories  = Accessories_Pool;
LinearCopySetup.CopyMode         = All;
\end{lstlisting}
Now the data transfer can be performed by a simple function call:
\lstset{escapechar=@,style=customc}
\begin{lstlisting}
ScanSystem.LinearSet(LinearCopySetup);
\end{lstlisting}
Observe that the above example copies the properties of $N_T$ number of systems from the beginning of the problem pool into the object. Since the number of systems can reside in the object is exactly $N_T$, the object is immediately filled-up in one transactions. Although the {\fontfamily{pcr}\selectfont \textcolor{blue}{LinearSet Configuration}} structure seems to be overcomplicated, it is necessary to keep the filling process flexible. During a longer simulation procedure, it maybe possible that only a portion of the systems in the object are need to be replaced from new ones from the pool. Figure\,\ref{CopyDataToObject} demonstrates how the linear copy function works. All the properties; that is, the two components of the time domains $t_i$, all the components of the initial conditions (initial state) $y_i$ and all the parameters $p_i$ of the $N_T$ number of systems are copied. The values of the accessories are omitted in Fig.\,\ref{CopyDataToObject} intentionally. Nevertheless, it is copied into the object even if its values are uninitialized in the pool. We shall see in Sec.\,\ref{details_initialization_finalization} that such an initialization can be easily done via a pre-declared device function. The memory allocation and the copy process is included into the program package to keep the code as general as possible. If external initialization of the accessories are necessary via the problem pool (as in the case of the time domain, the initial conditions and the parameters) the user have to keep in mind the data pattern discussed throughout Sec.\,\ref{problem_pool}. The required number of allocated double precision floating point numbers is $N_A \times N_T$.

The second possibility allows to copy several systems scattered in the problem pool via the {\fontfamily{pcr}\selectfont \textcolor{blue}{RandomSet()}} member function. The corresponding pre-defined structure for the copy set-up reads as
\lstset{escapechar=@,style=customc}
\begin{lstlisting}
struct RandomSetConfiguration
{
	int* IndicesInObject;
	int* IndicesInPool;
	int PoolSize;
	int NumberOfElements;
	
	double* PoolTimeDomain;
	double* PoolActualState;
	double* PoolParameter;
	double* PoolAccessories;
	
	CopyModePossibilities CopyMode;
};
\end{lstlisting}
The only difference from the {\fontfamily{pcr}\selectfont \textcolor{blue}{LinearSetConfiguration}} structure is that a full list of indices (in any order) for the serial numbers of the ODE systems have to be specified both for the problem pool and for the object. Therefore, memory allocations need to be performed first and the corresponding integer pointers must be passed to the {\fontfamily{pcr}\selectfont \textbf{\textcolor{green!40!black}{int}}}* type fields {\fontfamily{pcr}\selectfont \textcolor{blue}{IndicesInObject}} and {\fontfamily{pcr}\selectfont \textcolor{blue}{IndicesInPool}}. The number of the indices stored in the arrays must be exactly {\fontfamily{pcr}\selectfont \textcolor{blue}{NumberOfElements}}. The copy transaction can be performed similarly as in the firs case (linear set):
\lstset{escapechar=@,style=customc}
\begin{lstlisting}
ScanSystem.RandomSet(RandomCopySetup);
\end{lstlisting}
The example for the creation of an instance of the {\fontfamily{pcr}\selectfont \textcolor{blue}{RandomCopySetup}} structure and its fill-up with valid data is omitted here.

\subsection{Solver configuration and run} \label{solver_configuration}

Before running any kind of simulation, the configuration of the solver is also necessary. Again, a pre-defined structure
\lstset{escapechar=@,style=customc}
\begin{lstlisting}
struct SolverConfiguration
{
	int BlockSize;
	double InitialTimeStep;
	SolverAlgorithms Solver;
};
\end{lstlisting}
serves to perform this task. The main approach is that one thread in a GPU solves one independent system. The total number of launched threads $N_T$ defined in the object setup are organized into blocks of threads. The number of the threads in a block is prescribed by the {\fontfamily{pcr}\selectfont \textcolor{blue}{BlockSize}}. The detailed performance issues related to the number of the threads and the size of a block is presented in Sec.\,\ref{test_cases} for the three test cases. Their optimal values can depend on problem-to-problem, and also on the device architecture itself. Therefore, the interested reader is also referred to well-written textbooks \cite{Cheng2014,Soyata2018} for more details. As a rule-of-thumb, however, some general advice can be given for those who are new in GPU programming. It is important that the size of a block be the integer multiple of $32$ (size of a warp). Similarly, the number of the blocks in a single kernel run should be an integer multiple of the number of the streaming multiprocessors of a device. To harness the processing power of the GPU, the larger the number of the initiated blocks the better. The total number of the block can be calculated simply as $N_b=N_T/BlockSize$.

For the integration itself, two options can be chosen ({\fontfamily{pcr}\selectfont \textcolor{blue}{SolverAlgorithms}} field): {\fontfamily{pcr}\selectfont \textbf{\textcolor{blue}{RKCK45}}} (the adaptive Runge--Kutte--Kash--Carp method) and {\fontfamily{pcr}\selectfont \textbf{\textcolor{blue}{RK4}}} (the $4^{th}$ order Runge--Kutta scheme with fixed time step). In the solvers, there is no prediction for the initial time step. Thus, it has to be prescribed in the solver configuration structure via the {\fontfamily{pcr}\selectfont \textcolor{blue}{InitialTimeStep}} field. In case of the {\fontfamily{pcr}\selectfont \textbf{\textcolor{blue}{RK4}}} solver, this is also the fixed time step used during the integration. An example for a proper solver setup is
\lstset{escapechar=@,style=customc}
\begin{lstlisting}
SolverConfiguration SolverConfigurationSystem;

SolverConfigurationSystem.BlockSize       = 64;
SolverConfigurationSystem.InitialTimeStep = 1e-3;
SolverConfigurationSystem.Solver          = RKCK45;
\end{lstlisting}
with which the integration between the time instances $t_0$ and $t_1$ (different for each system) can be performed by simply call the solver member function of the object:
\lstset{escapechar=@,style=customc}
\begin{lstlisting}
ScanSystem.Solve(SolverConfigurationSystem);
\end{lstlisting}
In Sec.\,\ref{test_cases}, we shall demonstrate through several examples that how the solver member function can be called several times iteratively to produce e.g. bifurcation diagrams.

\subsection{The ODE function and ODE properties}

In the previous subsections, only the general preparations were introduced. The next task is to specify the problem via several device functions. Since function pointers cannot be passed to a CUDA kernel as an argument, these functions are pre-declared each with a specialized name with a well-defined purpose. This subsection focuses on the right-hand side of the ODE system and its properties. The following code snippet 
\lstset{escapechar=@,style=customc}
\begin{lstlisting}
__device__ void ParametricODE_Solver_OdeFunction(double* RightHandSide, int idx, int NoT, double t, double* StateVariable, double* Parameter)
{
	double y1 = StateVariable[idx + 0*NoT];
	double y2 = StateVariable[idx + 1*NoT];
	
	double p1 = Parameter[idx + 0*NoT];
	double p2 = Parameter[idx + 1*NoT];
	
	RightHandSide[idx + 0*NoT] = y2;
	RightHandSide[idx + 1*NoT] = y1 - y1*y1*y1 - p1*y2 + p2*cos(t);
}
\end{lstlisting}
shows an example of the right-hand side of the Duffing oscillator defined by Eqs.\,\eqref{duffing_equation_1}-\eqref{duffing_equation_2}. It must be given in the GPU device function {\fontfamily{pcr}\selectfont \textcolor{blue}{ParametricODE\_ Solver\_OdeFunction}} whose input arguments are also pre-defined and should not be modified. During the integration, each thread calls its own instance of this device function. Therefore, a unique thread identifier ({\fontfamily{pcr}\selectfont \textcolor{blue}{idx}}) is required to access data of a specific ODE system. With the help of the total number of threads ({\fontfamily{pcr}\selectfont \textcolor{blue}{NoT}}), the proper indices of the state variables and parameters of a system can be generated simply as $idx + i \cdot NoT$, where $i=0 \cdots N_{sys}$ for the state variables and $i=0 \cdots N_{par}$ for the parameters. Observe that in this way, coalesced global memory access can be achieved. For instance, the $32$ number $y_1$ values required by $32$ number of consecutive threads (consecutive thread identifier {\fontfamily{pcr}\selectfont \textcolor{blue}{idx}}) lye also consecutively in the global memory of a GPU. In order to keep the code clear, the state variables and the parameters are loaded into intermediate variables ($y_1-y_2$ and $p_1-p_2$ in this example) declared inside the ODE function. However, it is not a requirement. Naturally, the input parameter {\fontfamily{pcr}\selectfont \textcolor{blue}{t}} is the actual time instance of an integration process. The input arguments {\fontfamily{pcr}\selectfont \textcolor{blue}{StateVariable}}, {\fontfamily{pcr}\selectfont \textcolor{blue}{Parameter}} and {\fontfamily{pcr}\selectfont \textcolor{blue}{RightHandSide}} are self-explanatory; thus, their details are omitted here. With the above described technique, the ODE function can be given very similarly as in case of MATLAB; the last two lines in the code reflects the real physical content of the model. The only issue the user have to keep in mind is the aforementioned indexing, which necessary to distribute the independent ODE systems between the GPU threads.

The options to control the local error and the behaviour of the time step adaptation can be specified via the device function {\fontfamily{pcr}\selectfont \textcolor{blue}{ParametricODE\_ Solver\_OdeProperties}}:
\lstset{escapechar=@,style=customc}
\begin{lstlisting}
__device__ void ParametricODE_Solver_OdeProperties(double* RelativeTolerance, double* AbsoluteTolerance, double& MaximumTimeStep, double& MinimumTimeStep, double& TimeStepGrowLimit, double& TimeStepShrinkLimit)
{
	RelativeTolerance[0] = 1e-10;
	RelativeTolerance[1] = 1e-10;
	
	AbsoluteTolerance[0] = 1e-10;
	AbsoluteTolerance[1] = 1e-10;
	
	MaximumTimeStep     = 1.0e6;
	MinimumTimeStep     = 1.0e-12;
	TimeStepGrowLimit   = 5.0;
	TimeStepShrinkLimit = 0.1;
}
\end{lstlisting}
The number of the required relative and absolute error is the size of one system $N_{sys}$ (the indexing starts from zero). The maximum and minimum time steps are again self-explanatory. The variables {\fontfamily{pcr}\selectfont \textcolor{blue}{TimeStepGrowLimit}} and {\fontfamily{pcr}\selectfont \textcolor{blue}{TimeStepShrinkLimit}} control the maximum growth rate of the time step for an accepted step and the maximum shrink rate for a rejected step, respectively. Since all these variables are the same for all systems (shared by all threads), they are stored into the shared memory of the streaming multiprocessors, and kept them there until the end of an integration process. In case of the {\fontfamily{pcr}\selectfont \textbf{\textcolor{blue}{RK4}}} method with fixed time step, the above described options do not have any effect during the integration.

If the minimum time step size has been reached during a simulation, the solver tries to continue the integration with the prescribed minimum time step. Naturally, in this case, the tolerances shall not be maintained. If NaN values appears during the time step prediction phase, the next state is rejected and the updated time step is determined via the {\fontfamily{pcr}\selectfont \textcolor{blue}{TimeStepShrink Limit}} variable. If the minimum time step is reached with NaN values, the simulation will be stopped.

\subsection{The event function, event properties and event action function} \label{details_event_handling}

The device functions dedicated to event handling are presented through an example related to the dynamical model of the pressure relief valve introduced in Sec.\,\ref{model_pressure_relief_valve}. The user defined event functions given in an explicit form (see again Eqs.\,\eqref{UDEF1}-\eqref{UDEFn}) have to be implemented via the dedicated function {\fontfamily{pcr}\selectfont \textcolor{blue}{ParametricODE\_Solver\_EventFunction}} as:
\lstset{escapechar=@,style=customc}
\begin{lstlisting}
__device__ void ParametricODE_Solver_EventFunction(double* EventFunction, int idx, int NoT, double t, double* StateVariable, double* Parameter)
{
	double y1 = StateVariable[idx + 0*NoT];
	double y2 = StateVariable[idx + 1*NoT];
	
	EventFunction[idx + 0*NoT] = y2;
	EventFunction[idx + 1*NoT] = y1;
}
\end{lstlisting}
The indexing for the state variables (and parameters if needed) follows the convention discussed in more details in the previous section. Keep in mind that the serial number of a system $j$ for the variable {\fontfamily{pcr}\selectfont \textcolor{blue}{EventFunction}} in the indexing $idx + j \cdot NoT$ is between $0$ and $N_E-1$, where $N_E$ is the number of the defined event functions. In this specific example, the first event function (serial number $j=0$) means the detection of local extrema (maximum or minimum) of the variable $y_1$ (displacement of the valve), since $y_2$ is its velocity according to equation Eq.\,\eqref{pressure_relief_valve_1}. That is, the explicit form of the event function is $y_2=0$ specified again similarly as in case of MATLAB. The second function ($j=1$) detects events with vale position zero ($y_1=0$). The definition of the zero value of $y_1$ means closed valve; that is, the valve body is contacted with the seat. Therefore $y_1=0$ means non-smooth impact dynamics as well, see also Eqs.\,\eqref{impact_law_1}-\eqref{impact_law_3}.

The properties of each event function can be set by the following device function:
\lstset{escapechar=@,style=customc}
\begin{lstlisting}
__device__ void ParametricODE_Solver_EventProperties(int* EventDirection, double* EventTolerance, int* EventStopCondition, int& MaximumIterationForEquilibrium)
{
	EventDirection[0] = -1;
	EventDirection[1] = -1;
	EventTolerance[0] = 1e-6;
	EventTolerance[1] = 1e-6;
	
	EventStopCondition[0] = 1;
	EventStopCondition[1] = 0;
	
	MaximumIterationForEquilibrium = 50;
}
\end{lstlisting}
Here the {\fontfamily{pcr}\selectfont \textcolor{blue}{EventDirection}} variables follows the event handling rule of MATLAB. The value of $0$, $-1$ and $+1$ means detection for both directions, only negative direction and only positive direction, respectively. For instance, for the first event function $j=0$ ({\fontfamily{pcr}\selectfont \textcolor{blue}{EventDirection[0]}}), events are detected only for the local maxima of $y_1$ (its velocity changes its sign from negative to positive values). For each event function, an absolute tolerance {\fontfamily{pcr}\selectfont \textcolor{blue}{EventTolerance}} and a stop condition {\fontfamily{pcr}\selectfont \textcolor{blue}{EventStopCondition}} can be specified. The stop conditions is a serial number describing after how many detected events the solver have to stop the integration (zero means no stop). Observe that the number of the event functions and the aforementioned properties is $N_E$. This must be in accordance with the number already specified during the solver object initialization, see Sec.\,\ref{solver_object_initialization}. Finally, in case of autonomous systems, there may be fixed points sitting inside an event zone shown by the red dot in Fig.\,\ref{ConceptOfEventHandling}. If a trajectory is attracted by such a point, it never leave the event zone. The built in variable {\fontfamily{pcr}\selectfont \textcolor{blue}{MaximumIterationForEquilibrium}} is responsible for stopping the integration after the number of the accepted time steps inside an event zone reaches this value.

The main strength of the event handling system presented in this program package is the efficient managing of non-smooth dynamics (e.g. impact dynamics). In the above example, if the vale hits the seat ($y_1=0$) an impact law Eqs.\,\eqref{impact_law_1}-\eqref{impact_law_3} have to be applied to the dynamics. This can be easily included through the device function {\fontfamily{pcr}\selectfont \textcolor{blue}{ParametricODE\_Solver\_EventActionFunction}} called after every successfully event detection:
\lstset{escapechar=@,style=customc}
\begin{lstlisting}
__device__ void ParametricODE_Solver_EventActionFunction(int idx, int NoT, int EventIndex, int EventCounter, double t, double* StateVariable, double* Parameter)
{
	double y2 = StateVariable[idx + 1*NoT];
    
	double p5 = Parameter[idx + 4*NoT];
	
	if ( EventIndex == 1 )
	{
		StateVariable[idx + 1*NoT] = -p5 * y2;
	}
}
\end{lstlisting}
The function argument {\fontfamily{pcr}\selectfont \textcolor{blue}{EventIndex}} is served to distinguish the event functions, which is actually the serial number of the event function being detected. That is, the instructions inside the {\fontfamily{pcr}\selectfont \textbf{\textcolor{green!40!black}{if}}} statement are executed only for events detected by the second event function ($j=1$). The fifth parameter $p_5$ in this example is the coefficient of restitution $k<1$ describing kinetic energy loss during the impact. Observe that how the state variable $y_2$ changes during the function call. This variable is overwritten only via assigning a value to its original form {\fontfamily{pcr}\selectfont \textcolor{blue}{StateVariable[idx + 1*NoT]}} (keep in mind again the rule of the multi-thread indexing).

\subsection{The accessories function and the event accessories function} \label{details_accessories}

As it was already discussed in Sec.\,\ref{concept_accessories}, an ``ordinary'' accessory function is called after every successful time step during an integration. An example to extract the global maximum of the first component of a trajectory and its time instant is presented in the following listing:
\lstset{escapechar=@,style=customc}
\begin{lstlisting}
__device__ void ParametricODE_Solver_OrdinaryAccessories(double* Accessories, int idx, int NoT, double t, double* StateVariable, double* Parameter)
{
	double y1 = StateVariable[idx + 0*NoT];
	
	if ( y1>Accessories[idx + 0*NoT] )
	{
		Accessories[idx + 0*NoT] = y1;
		Accessories[idx + 1*NoT] = t;
	}
}
\end{lstlisting}
For this purpose, two accessory variables are necessary ($N_A=2$, see also Sec.\,\ref{solver_object_initialization}). The multi-thread indexing follows the same rule as before: $idx + k \cdot NoT$, where $k$ is the serial number of an accessory variable. The first one (with index $k=0$) stores the global maximum while the second one (with index $k=1$) register the time instance. The control logic inside the function {\fontfamily{pcr}\selectfont \textcolor{blue}{ParametricODE\_Solver\_OrdinaryAccessories}} is simple: if the actual value of the state variable $y_1$ is larger than the value stored in the corresponding accessory then overwrite it together with its time instance. Observe that the condition is checked only for the variable $y_1$, but both accessories are overwritten.

Although the evaluation of the ``ordinary'' accessory device function is fast, it cannot store e.g. the local maxima or minima of a trajectory. The synergy between the concept of accessories and event handling can help to overcome this difficulty. First, a proper event function have to be implemented to detect the local maxima (for details, see the previous section). Then, the {\fontfamily{pcr}\selectfont \textcolor{blue}{ParametricODE\_Solver\_EventAccessories}} device function called after every successful event detection can perform the storing process:
\lstset{escapechar=@,style=customc}
\begin{lstlisting}
__device__ void ParametricODE_Solver_EventAccessories(double* Accessories, int idx, int NoT, int EventIndex, int EventCounter, double t, double* StateVariable, double* Parameter)
{
	double y1 = StateVariable[idx + 0*NoT];
	
	bool EventCondition;
	EventCondition = (EventIndex==0)&&(EventCounter==3)
    
	if ((y1>Accessories[idx + 2*NoT])&&(EventCondition))
	{
		Accessories[idx + 2*NoT] = y1;
		Accessories[idx + 3*NoT] = t;
	}
}
\end{lstlisting}
The algorithm and the control flow is very similar to the ``ordinary'' accessories. The only difference is the inclusion of the serial number of the event and the serial number of the detection itself into the condition. The {\fontfamily{pcr}\selectfont \textcolor{blue}{EventCounter}} argument stores the number of the already detected events related to the event function with serial number {\fontfamily{pcr}\selectfont \textcolor{blue}{EventIndex}}. The above example stores the third local maximum and its time instance during an integration process into two additional accessories. Therefore, the value of $N_A$ have to be increased to four in this case. If the integration stops before the detection of three local maxima, the accessories will contain their initial values.

\subsection{The initialization and finalization functions} \label{details_initialization_finalization}

To properly detect the local maxima and minima of a trajectory, the proper initialization of the accessories is mandatory. This can be done using the {\fontfamily{pcr}\selectfont \textcolor{blue}{ParametricODE\_Solver\_Initialization}} device function as follows
\lstset{escapechar=@,style=customc}
\begin{lstlisting}
__device__ void ParametricODE_Solver_Initialization(int idx, int NoT, double t, double* TimeDomain, double* StateVariable, double* Parameter, double* Accessories)
{
	double y1 = StateVariable[idx + 0*NoT];
		
	Accessories[idx + 0*NoT] = y1;
	Accessories[idx + 1*NoT] = t;
	Accessories[idx + 2*NoT] = y1;
	Accessories[idx + 3*NoT] = t;
}
\end{lstlisting}
This function is called only once at the beginning of each integration phase. The proper initialization of the local extrema is the initial condition of the solution. Initializing the accessories corresponding to Fig.\,\ref{ConceptOfAccessories} by $y_1=10$, for instance, neither the global nor the local maxima can be detected as the initial values of the accessories remain always greater than any values of the trajectory.

The last issue the program code need to offer is the possibility to alter the state variables, accessories or even the time domain of the ODE systems at the end of the simulation. This can help to overcome some difficulties arise in certain problems. In case of the dual-frequency driven bubble model presented in Sec.\,\ref{model_keller_miksis}, the external forcing can be quasiperiodic. If one needs to perform iteration by sequentially call the {\fontfamily{pcr}\selectfont \textcolor{blue}{solver()}} member function of the ODE solver object, the proper track of the initial time is important. Otherwise, discontinuities are observed between every successive iterations. The solution to this problem is to set a very high value for the final time instance of the time domain, e.g. $t_1=10^6$. Prepare an event handling to stop the integration at specific properties of the solutions (e.g. local maximum of a component). Finally, through the {\fontfamily{pcr}\selectfont \textcolor{blue}{ParametricODE\_Solver\_Finalization}} dedicated device function, set the initial time instance $t_0$ to the time instance of simulation at the end:
\lstset{escapechar=@,style=customc}
\begin{lstlisting}
__device__ void ParametricODE_Solver_Finalization(int idx, int NoT, double t, double* TimeDomain, double* StateVariable, double* Parameter, double* Accessories)
{
	TimeDomain[idx + 0*NoT] = t;
}
\end{lstlisting}
This function is call only once at the end of every integration phase. In this way, the initial time can be properly tracked.

\subsection{Summary of the device functions}

The following device functions, their properties and behaviour were discussed thoroughly in this section:
\lstset{escapechar=@,style=customc}
\begin{lstlisting}
__device__ void ParametricODE_Solver_OdeFunction(double*, int, int, double, double*, double*);

__device__ void ParametricODE_Solver_OdeProperties(double*, double*, double&, double&, double&, double&);

__device__ void ParametricODE_Solver_EventFunction(double*, int, int, double, double*, double*);

__device__ void ParametricODE_Solver_EventActionFunction(int, int, int, int, double, double*, double*);

__device__ void ParametricODE_Solver_EventProperties(int*, double*, int*, int&);

__device__ void ParametricODE_Solver_OrdinaryAccessories(double*, int, int, double, double*, double*);

__device__ void ParametricODE_Solver_EventAccessories(double*, int, int, int, int, double, double*, double*);

__device__ void ParametricODE_Solver_Initialization(int, int, double, double*, double*, double*, double*);

__device__ void ParametricODE_Solver_Finalization(int, int, double, double*, double*, double*, double*);
\end{lstlisting}
They are the part of the {\fontfamily{pcr}\selectfont \textcolor{blue}{ParametricODESolver.cu}} and {\fontfamily{pcr}\selectfont \textcolor{blue}{Parametric ODESolver.cuh}} files. All the other definitions of the structures and classes and GPU kernel functions are defined here as well. Therefore, in order to use the program package it is necessary only to include the {\fontfamily{pcr}\selectfont \textcolor{blue}{ParametricODE Solver.cuh}} header file. The dedicated device functions to define the problem are placed at the top of the {\fontfamily{pcr}\selectfont \textcolor{blue}{.cu}} file to be able easily distinguish them from the rest of the code. This can be somewhat inconvenient if multiple models are intended to use. However, separating the above device functions into a different module, the speed of the code drops down approximately by $14\%$ (NVIDIA Titan Black card, NVCC version v7.5.17). The reason is that compiler optimization is usually much less efficient between modules, since each module must be complied separately and linking them take place only after the optimization. That is why we have kept all the device and kernel functions in the same module.

\subsection{Access to the simulated data}

One of the main concept of object oriented programming is to access the data of an object only through member functions. In this way, the access and the modification possibilities of the data elements are absolutely under the control of the member functions and thus by the object itself and its programmer. This requirement is fulfilled during the fill-up of the object from the problem pool via the {\fontfamily{pcr}\selectfont \textcolor{blue}{LinearSet()}} and {\fontfamily{pcr}\selectfont \textcolor{blue}{RandomSet()}} member functions.

After finishing a simulation, the computed data is immediately copied back the host (main memory of the CPU) from the device (global memory of the GPU). They are still reside, however, in the internal storages of the solver object, see again Sec.\,\ref{solver_object_initialization} and Sec.\,\ref{fill_the_solver_object}. The correct treatment would be to access the results via specialized member functions (e.g. {\fontfamily{pcr}\selectfont \textcolor{blue}{LinearGet()}} and {\fontfamily{pcr}\selectfont \textcolor{blue}{RandomGet()}}) and copy them back to the problem pool. This is an additional unnecessary memory transactions, since the data is already in the main memory of the CPU. In order to avoid this overhead, the internal storages have {\fontfamily{pcr}\selectfont \textbf{\textcolor{green!40!black}{public}}} qualifier, and the the data can be accessed directly though it violates the idea of data hiding in object oriented programming. Maybe in a later version of the program package, such member function shall be included.

The access to the variables time domain, actual state, parameters and accessories takes place via the pointers {\fontfamily{pcr}\selectfont \textcolor{blue}{h\_TimeDomain}}, {\fontfamily{pcr}\selectfont \textcolor{blue}{h\_ActualState}}, {\fontfamily{pcr}\selectfont \textcolor{blue}{h\_Parameters}} and
{\fontfamily{pcr}\selectfont \textcolor{blue}{h\_Accessories}}, respectively. The dereference indexing convention is the same as in case of the assembly of the problem pool and in the access of the variables in the pre-declared device functions:
\lstset{escapechar=@,style=customc}
\begin{lstlisting}
h_TimeDomain [idx + i*NoT];
h_ActualState[idx + j*NoT];
h_Parameters [idx + k*NoT];
h_Accessories[idx + l*NoT];
\end{lstlisting}
Here the variable {\fontfamily{pcr}\selectfont \textcolor{blue}{idx}} is the serial number of the system, its value is between $0$ and $N_T-1$. The indices $i\in[0, 1]$, $j=0 \cdots N_{sys}-1$, $k=0 \cdots N_{par}-1$ and $l=0 \cdots N_A-1$ are the serial numbers of the components of the variables.


\section{Discussion: test cases, performances and profiling} \label{test_cases}

The main aim of this section---besides presenting some results through diagrams---is to demonstrate the efficiency of the code. Instead of the direct comparison of the runtime between CPUs and GPUs which is a common test process \cite{Murray2012,Stone2014,Rodriguez2015}, the utilization of the streaming multiprocessors (SMs) and arithmetic function units are presented. They are obtained via the NVIDIA Visual Profiler (release 7.5). Besides the guided visual profiling, several metrics and events are collected directly in order to obtain a detailed insight about the performances of the kernel functions.

Although comparing CPUs with GPUs is important and interesting, they gave no information on how efficient the GPU code is, and how much of the theoretical processing power is harnessed. That is why this section focuses only on the aforementioned analyses of the GPU kernels which are usually missing in the recent studies dealing with the integration of large number of independent ODE systems. If a code is poorly written and only a portion of the SMs are well utilized, then comparing such a code with implementations on other architectures are meaningless. In case of CPUs, the compiler can do much work to optimize the code even if it is suboptimal. In case of GPUs, however, much more confident knowledge on the underlying architecture and memory hierarchy is necessary from the programmer to write efficient codes. In the massively parallel nature of GPUs, if the organization of the threads is poor and their workload is unbalanced then the compiler can barely compensate the conceptual mistakes.

Before getting involved into the details, a preliminary discussion is necessary about the general factors in the performance of a GPU kernel function. First, every global memory access must be coalesced to maximise the memory throughput. Efficient memory transactions are mandatory to feed the fast GPU cores with enough data. The data patter of the problem pool and the internal storages of the solver object fulfil this requirement. Therefore, in case of no thread divergence (e.g. with the RK4 solver), all the global memory load and store efficiency is $100\%$.

Second, if it is possible, load some data to the shared memory which are common for threads reside in the same thread block. Shared memory of an SM is much faster than global memory, but its size is limited only to 48\,Kb to 16\,Kb (programmable). In the present program package, shareable variables are the ones set by the ODE and event option pre-declared device functions (e.g. the tolerances). They shared by all the threads in a block, but their overall size is limited to few bytes. Therefore, even though they are loaded into the shared memory to ease the pressure on global memory, they cannot provide a solid solution for the global memory access.

The general concept to hide global memory latency in a GPU is to reside hundreds or even thousands of threads simultaneously on an SM. They are organized into blocks of threads, and in each block the threads are organized into warps (32 threads). An instructions is performed on a warp only if all the necessary data is arrived from the global memory. The more the number of the residing warps, the more the chance an SM can find a warp eligible to perform an instruction. Naturally, there are limitations on how many threads, blocks, and warps can reside in an SM: a maximum of $2048$ threads, $16$ blocks and $64$ warps on the Kepler architecture (used here). Moreover, these values are further limited by the usage of the SM memory resources as well. Fast switching between warps (context switch) by the SM is possible due to the available large registry file: every operations take place through registers, and thus every required data in an instruction must be in the registry. Therefore, every thread must have their own portion from the registry to store their intermediate data and results, and to be able to perform fast context switch if necessary. Registers are the fastest memories in an SM, but the size is again limited to $65536$ number of $32$-bit registers. With a quick calculation, if one intends to assign $2048$ number of threads to an SM, then the number of the registers allocated to a thread must be no more than $32=65536/2048$. The upper limit of the register usage can be set as a compiler option \textit{-maxrregcount=x}, where \textit{x} is an integer value limited by compute capability of an architecture ($255$ in Kepler). If the maximum number of registers per threads is set, for instance, to $64$, the maximum number of residing threads is $1024$ instead of the absolute limit of $2048$.

Now the concept of occupancy $O_{SM}$ can be easily defined as the ratio of the actual number of residing threads in an SM and the theoretical limit. In the above example: $O_{SM}=1024/2048=0.5=50\%$. The less the occupancy is the less is the theoretical variability of an SM to choose between warps. On the other hand, in case of a registry ``hungry'' kernel, the SM have to switch the context unnecessarily many times or wait for data from the global memory if the allocated number of registers per threads are limited only to $32$. This can also degrade the performance of a kernel, and thus the pursuit for $100\%$ occupancy is not an ultimate goal.

Although the register pressure is usually the main factor for the upper limit of the theoretical occupancy, one has to keep in mind also the above mentioned limits for the number of block and warps on an SM. In the next subsections, detailed performance analysis will be given taking into account the aforementioned factors.

\subsection{Test case: Duffing equation} \label{test_duffing}

The simplest test cases are related to the non-stiff Duffing system described by Sys.\,\eqref{duffing_equation_1}-\eqref{duffing_equation_2} and Sys.\,\eqref{duffing_equation_1}-\eqref{duffing_equation_4} in case of the computation of the Lyapunov exponents. Therefore, these are perfect examples to compare the performances of the different solvers and to test whether the code is memory bandwidth or computation throughput limited for such simple systems. The ideal case is when the arithmetic function units of the GPU is fully utilized; that is, the code is computation throughput limited. This is harder to achieve if the evaluation of the right hand side functions are computationally less intensive, which shifts the pressure towards the global memory transactions.

\subsubsection{Computation of Poincar\'e sections}

\begin{figure}[ht!]   
	\centering
		\includegraphics[width=8.6cm]{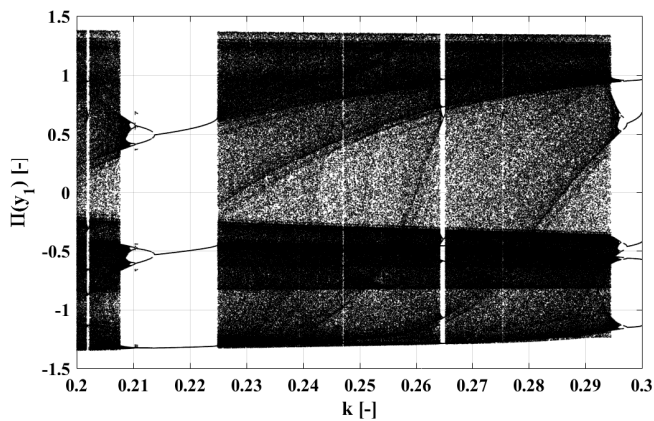}
	\caption{Bifurcation diagram of the Duffing oscillator; that is, the first component of the Poincar\'e section $\Pi(y_1)$ as a function of the damping parameter $k$. The amplitude of the excitation is $B=0.3$, the stiffens of the beam is $\delta=1$ and the excitation frequency is $\omega=1$. The number of the threads $N_T=30720$ that is also the number of the employed control parameters $k$ distributed equally.}
	\label{Duffing1}
\end{figure}

The first test case computes a simple bifurcation diagram where the first component of the Poincar\'e section $\Pi(y_1)$ is plotted against the control parameter $k$ (damping factor), see Fig.\,\ref{Duffing1}. The $N_T$ number of $k$ values are distributed equally between $0.2$ and $0.3$. Since the period of the excitation is $2 \pi$, the period of the state space in time is exactly $2 \pi$ as well. Therefore, the points of the Poincare\'e section can be obtained by sampling the continuous trajectory by $2 \pi$. By setting the time domain to $t_0=0$ and $t_1=2 \pi$ for every thread, the points can be calculated via a simple cycle:
\lstset{escapechar=@,style=customc}
\begin{lstlisting}
for (int i=0; i<NumberOfIterations; i++)
{
	ScanSystem.Solve(SolverConfigurationSystem);
	// Save the Poincare sections
	...
}
\end{lstlisting}
Observe that one needs only the endpoints of each integration. After every integration, the endpoints will be the new values of the internal state variables both on the device (GPU) and in the host (CPU). Therefore, no additional effort is necessary to initialize the system inside the loop. The entire program code can be found in the supplementary material Duffing\_v1.zip. After compiling via the prepared makefile and running the resulted executable, the code creates the datafile Duffing\_v1.txt, in which the the two parameters (first and second columns) and the two components of the Poincar\'e sections $\Pi(y_1)$ and $\Pi(y_2)$ (third and fourth columns) are stored. The results of the first $1024$ number of iterations are regarded as initial transients and are discarded, and only the next $32$ number of points are saved. This first test case will be referred to as Duffing1 in the rest of this paper.

In the following, let us analyse thoroughly the performance characteristics of the kernel functions via the NVDIA Visual Profiler. The key factors have significant effect are the maximum number of registers per threads $N_{reg}$ set by the compiler flag \textit{-maxrregcount=$N_{reg}$}; the number of threads in a single block $N_{t/b}$ specified in the structure {\fontfamily{pcr}\selectfont \textcolor{blue}{SolverConfigurationSystem}} (see also Sec.\,\ref{solver_configuration}); and finally the total number of active threads during a single run $N_T$ given in the structure {\fontfamily{pcr}\selectfont \textcolor{blue}{ConfigurationSystem}} (see also Sec.\,\ref{solver_object_initialization}).

One of the most important quantity of a kernel function is its theoretical occupancy $O_{TH}$ calculated by the CUDA Occupancy Calculator \footnote{https://developer.download.nvidia.com/compute/cuda/CUDA\_Occupancy\_calculator.xls}. This in general depends on the compute capability of the hardware (3.5 for our Kepler architecture), the available and used amount of shared memory (not important in our cases due to the required few bytes of shared memory), the number of the used registers per threads and the number of threads in a block. The latter two are the most relevant factors. Keep in mind that this occupancy is only a theoretical one, the achieved occupancy of the SMs $O_{SM}$ is depend on many other factors (e.g. the total number of launched threads $N_T$) and can be obtained via profiling.

Another factor play a significant role in the performance and runtime is the average number of blocks assigned to an SM. During a kernel run, CUDA tries to distribute the total number of blocks $N_b$ equally among the SMs. Therefore, if the total number of blocks per SM $N_{b/SM}=N_{T}/N_{t/b}/N_{SM}$ is not an integer number then at the final phase of a simulation, some SMs shall be idle due to the lack of the number of the remaining blocks. Here $N_{SM}=15$ is the number of the streaming multiprocessors of our GPU. Practically, this means that the total number of threads $N_T$ must be set precisely to maximise the performance.

Beside the computational time of a kernel function normalized to a single thread $t_{c/t}$, the number of the eligible warps per active cycle $EW$ of an SM, the arithmetic function utilization $AFU$, the FLOP efficiency $FE$ and the multiprocessor activity $MA$ are profiled and their data are collected. The latter three are the key measures to monitor how efficiently the hardware computational resources are harnessed.

The profiling results as a function of $N_{reg}$, $N_{t/b}$ and $N_T$ for the kernel function corresponding to the fourth order Runge--Kutta (RK4) solver with fixed time step ($10^{-2}$) are summarized in Tab.\,\ref{tab:nvvp_duffing_v1_RK4}. The registers required to avoid spilling is $62$ (turns out at compile time). Therefore, the maximum number of registers used here is $64$. Increasing it further would result in the decrease of the theoretical occupancy without any further benefit.

\begin{table}
\caption{Summary of the profiling results of the test case Duffing1 corresponding to the kernel function of the fourth order Runge--Kutta solver using a fixed time step of $10^{-2}$.}
\label{tab:nvvp_duffing_v1_RK4}
\begin{tabular}{ccccccccccc}
\hline\noalign{\smallskip}
$N_{reg}$ & $N_{t/b}$ & $N_T$ & $O_{TH}$ & $O_{SM}$ & $N_{b/SM}$ & $t_{c/t}$ & $EW$ & $AFU$ & $FE$ & $MA$ \\
          &           &       & (\%)     & (\%)     &            & ($\mu s$) &      &       & (\%) & (\%) \\
\noalign{\smallskip}\hline\noalign{\smallskip}
64 & 64  & 7680  & 50  & 24.9 & 8   & 1.16 & 2.80  & 10 & 53.3 & 98.5 \\
64 & 128 & 7680  & 50  & 24.9 & 4   & 1.11 & 2.81  & 10 & 53.3 & 98.6 \\
40 & 96  & 7680  & 75  & 24.1 & 5.3 & 1.37 & 2.58  & 9  & 43.2 & 89.3 \\
40 & 128 & 7680  & 75  & 24.9 & 4   & 1.02 & 2.63  & 9  & 50.0 & 98.4 \\
40 & 192 & 7680  & 75  & 23.7 & 3   & 1.27 & 2.54  & 8  & 41.7 & 90.4 \\
32 & 128 & 7680  & 100 & 24.9 & 4   & 1.21 & 2.52  & 9  & 47.3 & 98.4 \\
\noalign{\smallskip}\hline
64 & 64  & 11520 & 50  & 37.4 & 12  & 0.86 & 4.79  & 10 & 57.0 & 98.6 \\
64 & 128 & 11520 & 50  & 37.4 & 6   & 0.99 & 4.79  & 10 & 57.0 & 98.6 \\
40 & 96  & 11520 & 75  & 37.4 & 8   & 1.02 & 4.53  & 10 & 55.6 & 98.7 \\
40 & 128 & 11520 & 75  & 37.4 & 6   & 1.04 & 4.53  & 10 & 55.5 & 98.5 \\
40 & 192 & 11520 & 75  & 37.2 & 4   & 1.05 & 4.51  & 10 & 55.0 & 98.5 \\
32 & 128 & 11520 & 100 & 37.4 & 6   & 1.08 & 4.37  & 10 & 52.8 & 98.5 \\
\noalign{\smallskip}\hline
64 & 64  & 7680  & 50  & 24.9 & 8   & 1.16 & 2.80  & 10 & 53.3 & 98.5 \\
64 & 64  & 11520 & 50  & 37.4 & 12  & 0.86 & 4.79  & 10 & 57.0 & 98.6 \\
64 & 64  & 15360 & 50  & 49.8 & 16  & 1.01 & 6.70  & 10 & 57.3 & 98.6 \\
64 & 64  & 30720 & 50  & 49.8 & 32  & 0.94 & 6.70  & 10 & 57.6 & 99.0 \\
64 & 64  & 61440 & 50  & 49.8 & 64  & 0.90 & 6.70  & 10 & 57.6 & 99.1 \\
\noalign{\smallskip}\hline
32 & 128 & 7680  & 100 & 24.9 & 4   & 1.21 & 2.51  & 9  & 47.3 & 98.4 \\
32 & 128 & 11520 & 100 & 37.4 & 6   & 1.08 & 4.37  & 10 & 52.8 & 98.5 \\
32 & 128 & 15360 & 100 & 49.8 & 8   & 1.07 & 6.18  & 10 & 53.6 & 98.6 \\
32 & 128 & 30720 & 100 & 99.8 & 16  & 1.01 & 12.68 & 10 & 54.1 & 99.0 \\
32 & 128 & 61440 & 100 & 99.7 & 32  & 0.97 & 12.67 & 10 & 54.4 & 99.4 \\
\noalign{\smallskip}\hline
\end{tabular}
\end{table}

The results shows that there is no significant difference between the configurations in terms of the runtime $t_{c/t}$ (except some special cases discussed later). As a first note, the calculation of the theoretical occupancy is based only on $N_{reg}$ and $N_{t/b}$. However, in order to get close to this theoretical value in practice, the total number of threads $N_T$ must be sufficiently high. For instance, if the maximum possible number of threads in our SM is $2048$ then the required number of total threads is $N_T=2048*15=30720$ in order to achieve nearly $100\%$ occupancy $O_{SM}$. This is clearly shown in Tab.\,\ref{tab:nvvp_duffing_v1_RK4}. Observe also that the number of the eligible warps per active cycle $EW$ correlates very well with $O_{SM}$.

The arithmetic function utilization $AFU$ (measured as an integer level between $0$ and $10$) and the multiprocessor activity $MA$ is very high in almost every cases (approximately if $N_T \geq 11520$). Although $AFU=10$ and $MA>98\%$, the FLOP efficiency $FE$ seems to be saturated at $57.6\%$. This may due to the required mixed integer (e.g. index and memory address calculations) and floating point instructions. According to the runtime and the value of $FE$, the optimal configuration for this problem is when $N_{reg}=64$ and $N_{t/b}=64$ with $N_T \geq 11520$.

The very bad performance shown in the third row of Tab.\,\ref{tab:nvvp_duffing_v1_RK4} is due to the non-integer number of average block per SM. This is clear also from the low multiprocessor activity $MA<90\%$. The other relatively high runtime with $N_T=7680$ is due to the low achieved occupancy $O_{SM}<25\%$. Therefore, an SM can choose between very few eligible warps during a computation. Nevertheless, with high enough block per SM  combined with high number of registers (first two row), the GPU utilization can still be high. That is, the code is efficient even employing very simple systems and very few number of threads.

\begin{table}
\caption{Summary of the profiling results of the test case Duffing1 corresponding to the kernel function of the Runge--Kutta--Cash--Karp solver. Both the absolute and relative tolerances of the adaptive time stepping are $10^{-9}$.}
\label{tab:nvvp_duffing_v1_RKCK45}
\begin{tabular}{ccccccccccc}
\hline\noalign{\smallskip}
$N_{reg}$ & $N_{t/b}$ & $N_T$ & $O_{TH}$ & $O_{SM}$ & $N_{b/SM}$ & $t_{c/t}$ & EW & AFU & FE   & MA  \\
          &           &       & (\%)     & (\%)     &            & ($\mu s$) &    &     & (\%) & (\%)\\
\noalign{\smallskip}\hline\noalign{\smallskip}
64  & 64 & 7680  & 50 & 22.0 & 8   & 0.64 & 2.40 & 8  & 38.0 & 95.3 \\
64  & 64 & 11520 & 50 & 33.3 & 12  & 0.57 & 4.10 & 9  & 42.4 & 95.8 \\
64  & 64 & 15360 & 50 & 44.8 & 16  & 0.55 & 5.91 & 9  & 44.1 & 96.2 \\
64  & 64 & 30720 & 50 & 45.9 & 32  & 0.44 & 6.04 & 10 & 46.0 & 97.1 \\
64  & 64 & 61440 & 50 & 46.6 & 64  & 0.46 & 6.15 & 10 & 48.4 & 98.1 \\
\noalign{\smallskip}\hline
128 & 64 & 7680  & 25 & 22.0 & 8   & 0.62 & 2.38 & 8  & 38.9 & 95.3 \\
128 & 64 & 11520 & 25 & 19.3 & 12  & 0.70 & 2.02 & 8  & 35.1 & 93.0 \\
128 & 64 & 15360 & 25 & 23.1 & 16  & 0.57 & 2.49 & 9  & 42.1 & 97.4 \\
128 & 64 & 30720 & 25 & 23.6 & 32  & 0.52 & 2.54 & 9  & 43.9 & 98.1 \\
128 & 64 & 61440 & 25 & 23.9 & 64  & 0.48 & 2.58 & 9  & 45.6 & 98.7 \\
\noalign{\smallskip}\hline
\end{tabular}
\end{table}

The profiling results corresponding to the kernel function of the Runge--Kutta--Cash--Karp (RKCK45) solver is summarized in Tab.\,\ref{tab:nvvp_duffing_v1_RKCK45}. Although, a single step with the RKCK45 algorithm is much more computationally expensive than with the simple RK4 scheme, the overall simulation time dropped down by almost a factor of two due to the much less number of time steps. Here the required number of registers to avoid spilling is $111$. Thus, configurations with $128$ number of maximum registers are also included. The best configurations are again correspond to $N_{reg}=64$ and $N_{t/b}=64$, though the total number of the threads have to be increased to $30720$ in order to fully utilize the GPU.

Because of the adaptive time step, the simulation time for each thread can be different according to the overall required time steps and the number of the rejected steps. This results in the well-know thread divergence phenomenon. Since every thread in a warp must finish their work exactly at the same time, the total simulation time of a warp is determined by its slowest thread (the threads already finished their work become idle). If some threads require much larger number of steps then the others, the utilization of the GPU can be very unbalanced. The effect of the thread divergence phenomenon is the somewhat lower multiprocessor activity $MA$ shown in Tab.\,\ref{tab:nvvp_duffing_v1_RKCK45}. It is still, however, higher than $97\%$ in the optimal configurations. The system is non-stiff and the solutions are ``smooth'' enough; therefore, the computational requirements for each system is nearly the same.

\subsubsection{Computation of the maximum of $y_1$ via accessories and event handling}

\begin{figure}[ht!]   
	\centering
		\includegraphics[width=8.6cm]{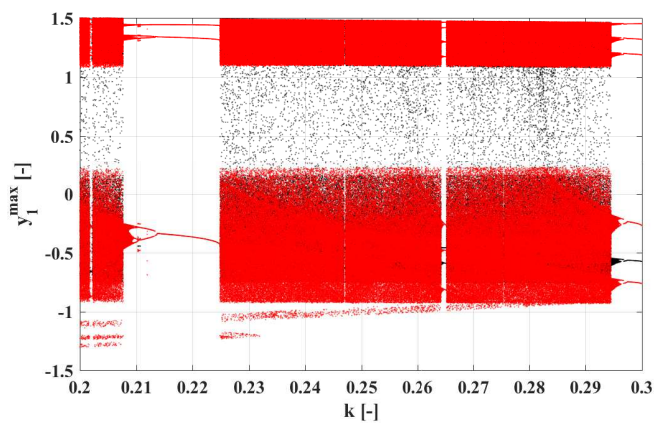}
	\caption{The maximum of the first component of the solution $y_1^{max}$ registered at every consecutive iterations as a function of the damping parameter $k$. The black and red dots correspond to the results obtained via the accessories and event handling, respectively. The amplitude of the excitation is $B=0.3$, the stiffens of the beam is $\delta=1$ and the excitation frequency is $\omega=1$. The number of the threads $N_T=30720$ that is also the number of the employed control parameters $k$ distributed equally.}
	\label{Duffing23}
\end{figure}

The next two test cases calculate the maxima of the first component of the solutions $y_1^{max}$ of the consecutive iterations as a function of the damping rate $k$ as control parameter. The results are depicted by Fig.\,\ref{Duffing23} where the black and red points are obtained via ordinary accessories (referred to as Duffing2) and event handling (referred to as Duffing3), respectively. The corresponding supplementary materials are Duffing\_v2.zip and Duffing\_v3.zip. It must be emphasized that in case of a long trajectory, the maxima of the successive integrations can be different (e.g. in case of chaotic solution) resulted in scattered points in Fig.\,\ref{Duffing23} at some parameter values.

The main aim of this section is to investigate the effect of additional control logic on the performance of the code. As a reminder, using the accessories to store the maxima of every integration needs only a comparison and a load/store transaction. In case of event handling, iterations of the precise event detection is also required. Moreover, event handling actually detects the local maxima (if it exists since the function can be monotonic) whereas accessories always detects the global maxima during a single integration process. Therefore, the two approach are not exactly equivalent, though the absolute maxima (envelopes in Fig.\,\ref{Duffing23}) of a long trajectory are the same. Such a difference can be seen, for instance, in the lower part of the figure near $k=0.3$.

The definition of the sole event function as $F_{E1}=y_2$ ($y_2=0$) means the detection of the extrema points of $y_1$ (since $y_2=\dot{y}_1$). By setting the event direction to $-1$, only the local maxima are detected. The tolerance and the stop condition are set to $10^{-6}$ and $0$ (continue after detection), respectively.

Based on the results of the previous section and to shorten the discussion, only the kernel function of the RKCK45 solver and the combination $N_{reg}/N_{t/b}=64/64$ is investigated. The profiled data is summarized in Tab.\,\ref{tab:nvvp_duffing_v23_RKCK45} for both test cases (accessories and event handling). Interestingly, the additional control logic has marginal effect on the runtime and the utilization of the SMs, the differences are buried by the deviation of measured values (compare also with the third, fourth and fifth rows of Tab.\,\ref{tab:nvvp_duffing_v1_RKCK45}). Therefore, the implementation of the accessories and the event handling is very efficient, at least if they are used with care (low number of accessories and event functions). We shall see in Sec.\,\ref{test_impact_dynamics} that using multiple event functions can degrade the utilization of the arithmetic function units of the GPU significantly. Parenthetically, it should be noted that the required number of registers to avoid spilling are $128$ and $134$ for the test cases Duffing2 and Duffing3, respectively.

\begin{table}
\caption{Summary of the profiling results of the test cases Duffing2 and Duffing3 corresponding to the kernel function of the Runge--Kutta--Cash--Karp solver. Both the absolute and relative tolerances of the adaptive time stepping are $10^{-9}$.}
\label{tab:nvvp_duffing_v23_RKCK45}
\begin{tabular}{ccccccccccc}
\hline\noalign{\smallskip}
$N_{reg}$ & $N_{t/b}$ & $N_T$ & $O_{TH}$ & $O_{SM}$ & $N_{b/SM}$ & $t_{c/t}$ & EW & AFU & FE   & MA  \\
          &           &       & (\%)     & (\%)     &            & ($\mu s$) &    &     & (\%) & (\%)\\
\noalign{\smallskip}\hline\noalign{\smallskip}
\multicolumn{3}{c}{Accessories:} \\
64 & 64 & 15360  & 50 & 44.8 & 16   & 0.54 & 5.92 & 9  & 43.9 & 96.3 \\
64 & 64 & 30720  & 50 & 45.9 & 32   & 0.51 & 6.05 & 10 & 45.9 & 97.2 \\
64 & 64 & 61440  & 50 & 47.0 & 64   & 0.43 & 6.23 & 10 & 48.8 & 98.2 \\
\noalign{\smallskip}\hline
\multicolumn{3}{c}{Event handling:} \\
64 & 64 & 30720  & 50 & 46.1 & 32   & 0.52 & 6.00 & 10 & 45.1 & 97.0 \\
64 & 64 & 61440  & 50 & 46.9 & 64   & 0.46 & 6.11 & 10 & 47.4 & 98.1 \\
64 & 64 & 122880 & 50 & 47.1 & 128  & 0.45 & 6.17 & 10 & 48.5 & 98.6 \\
\noalign{\smallskip}\hline
\end{tabular}
\end{table}

\subsubsection{Computation of the Lyapunov exponent}

\begin{figure}[ht!]   
	\centering
		\includegraphics[width=8.6cm]{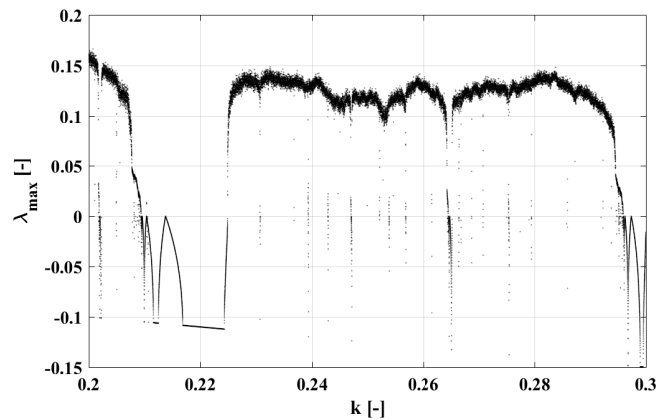}
	\caption{The largest Lyapunov exponent $\lambda_{max}$ as a function of the damping parameter $k$. The amplitude of the excitation is $B=0.3$, the stiffens of the beam is $\delta=1$ and the excitation frequency is $\omega=1$. The number of the threads $N_T=30720$ that is also the number of the employed control parameters $k$ distributed equally.}
	\label{Duffing4}
\end{figure}

The last test case related to the Duffing oscillator is the computation of its Lyapunov exponent at the parameter sets presented already in Fig.\,\ref{Duffing1} and in Fig.\,\ref{Duffing23} (supplementary material Duffing\_v4.zip). The results are shown in Fig.\,\ref{Duffing4} which is in good accordance with the previous results. That is, the Lyapunov exponent is positive at parameter values where chaos is presented (scattered points in Figs.\,\ref{Duffing1}-\ref{Duffing23}) while it is negative for periodic solutions (finite number of points in Fig.\,\ref{Duffing1}-\ref{Duffing23}). At bifurcation points, the largest Lyapunov exponent is zero.

\begin{table}
\caption{Summary of the profiling results of the test cases Duffing4 corresponding to the kernel function of the Runge--Kutta--Cash--Karp solver. Both the absolute and relative tolerances of the adaptive time stepping are $10^{-9}$.}
\label{tab:nvvp_duffing_v4_RKCK45}
\begin{tabular}{ccccccccccc}
\hline\noalign{\smallskip}
$N_{reg}$ & $N_{t/b}$ & $N_T$ & $O_{TH}$ & $O_{SM}$ & $N_{b/SM}$ & $t_{c/t}$ & EW & AFU & FE   & MA  \\
          &           &       & (\%)     & (\%)     &            & ($\mu s$) &    &     & (\%) & (\%)\\
\noalign{\smallskip}\hline\noalign{\smallskip}
64 & 64 & 15360  & 50 & 45.5 & 16 & 1.30 & 4.91 & 9  & 43.6 & 96.6 \\
64 & 64 & 30720  & 50 & 46.6 & 32 & 1.17 & 5.05 & 10 & 46.1 & 97.5 \\
64 & 64 & 61440  & 50 & 47.3 & 64 & 1.08 & 5.67 & 10 & 48.4 & 98.3 \\
\noalign{\smallskip}\hline
\end{tabular}
\end{table}

Table\,\ref{tab:nvvp_duffing_v4_RKCK45} summarizes the profiled results employing the first three kernel configurations shown in Tab.\,\ref{tab:nvvp_duffing_v23_RKCK45} (Accessories part). The utilization of the arithmetic function units $AFU$ and the multiprocessor activity $MA$ is still high, there is no substantial difference compared to the previously profiled values. The computation time $t_{c/t}$, on the other hand, is increased drastically (more than a factor of two). The reason is two fold: a) The system size is increased from two to four. This alone means more core intensive problem. Since the processing power is already highly utilized, adding more equations must definitely resulted in the increase of the runtime. b) The linearized system Eqs.\,\eqref{duffing_equation_3}-\eqref{duffing_equation_4} in polar coordinates contains many $\sin$ and $\cos$ transcendental functions. They are very expensive calculations (actually both on CPU and GPU). This is the second factor which can increase the runtime significantly. Naturally, as the following code snippet shows, the two-kinds of trigonometric functions are pre-computed before use softening the pressure on the arithmetic units of the GPU:
\lstset{escapechar=@,style=customc}
\begin{lstlisting}
__device__ void ParametricODE_Solver_OdeFunction(double* RightHandSide, int idx, int NoT, double t, double* StateVariable, double* Parameter)
{
	double y1 = StateVariable[idx + 0*NoT];
	double y2 = StateVariable[idx + 1*NoT];
	double y3 = StateVariable[idx + 2*NoT];
	double y4 = StateVariable[idx + 3*NoT];
	
	double p1 = Parameter[idx + 0*NoT];
	double p2 = Parameter[idx + 1*NoT];
	
	RightHandSide[idx + 0*NoT] = y2;
	RightHandSide[idx + 1*NoT] = y1 - y1*y1*y1 - p1*y2 + p2*cos(t);
	
	double g1 = 1 - 3*y1*y1;
	double g2 = -p1; 
	
	double s;
	double c;
	sincos(y4, &s, &c);
	
	RightHandSide[idx + 2*NoT] = y3*((1.0+g1)*s*c+g2*s*s);
	RightHandSide[idx + 3*NoT] = -s*s+(g1*c+g2*s)*c;
}
\end{lstlisting}
Inspecting the increase in the complexity of the right hand side of the system with the linearized part, the magnitude of the increase in the runtime is reasonable. The required number of registers to avoid spilling is $208$.

\subsection{Test case: Keller--Miksis equation (bubble model)} \label{test_keller_miksis}

\begin{figure}[ht!]   
	\centering
		\includegraphics[width=8.6cm]{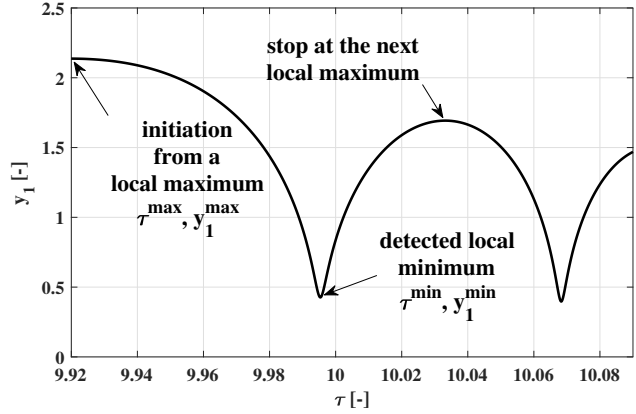}
	\caption{Demonstration of a bubble collapse via a dimensionless bubble radius vs. time curve. An integration start from a local maximum $y_1^{max}$ at time instant $\tau^{max}$ and ends at the next local maximum. During the integration, the local minimum $y_1^{min}$ and its time instant $\tau^{min}$ is also determined.}
	\label{BubbleCollapse}
\end{figure}

The following test case calculates the collapse strength of an air filled single spherical bubble placed in liquid water and subjected to dual-frequency ultrasonic irradiation. For the details of the mathematical modelling describing the radial pulsation of such a bubble, the reader is referred back to Sec.\,\ref{model_keller_miksis}. An example for the radial oscillation of a bubble is demonstrated in Fig.\,\ref{BubbleCollapse}, in which the dimensionless bubble radius $y_1=R(t)/R_E$ is presented as a function of the dimensionless time $\tau$. Keep in mind again that $R_E=10 \mathrm{\mu m}$ is the equilibrium bubble radius of the unexcited system. Parenthetically, function in Fig.\,\ref{ConceptOfAccessories} corresponds to the same trajectory taken at a different time interval. At certain parameter values, the oscillation can be so violent that at the minimum bubble radius the temperature can exceed several thousands of degrees of Kelvin initiating even chemical reactions \cite{Yasui2008,Stricker2014}. This phenomenon is called the collapse of a bubble. In the literature, there are various quantities characterising the strength of the collapse which is the keen interest of sonochemistry \cite{Mettin2015}. For instance, the relative expansion $y_{exp}=(R_{max}-R_E)/R_E = y_1^{max}-1$ \cite{Rossello2016} or the compression ratio $y_1^{max}/y_1^{min}$ \cite{Mettin2015} are good candidates. In this sense, a bubble collapse can be characterized by the radii of a local maximum $y_1^{max}$ and the subsequent local minimum $y_1^{min}$, see also Fig.\,\ref{BubbleCollapse}. Observe that the time scales can be very different near the local maximum and the local minimum, compare also again with $y_1^{min3}$ in Fig.\,\ref{ConceptOfAccessories}.

In general, during a long term oscillation of a bubble, the collapse strengths are different from collapse to collapse (e.g. due to chaotic behaviour or quasiperiodic forcing). Therefore, the properties of multiple collapses are registered in order to obtain a realistic picture about the bubble behaviour. The easiest way to do this is to integrate the system from a local maximum to the next local maximum (one iteration) meanwhile monitoring and detecting the local minimum. The iteration process can be repeated arbitrary many times. One iteration means the call of the solver member function: {\fontfamily{pcr}\selectfont \textcolor{blue}{ScanSystem.Solve(SolverConfigurationSystem)}}. After each iteration, the collapse properties are saved. It must be noted that some researchers take into account the elapsed time during a collapse \cite{Kanthale2007}. Thus, the time instances of the maxima and minima are stored as well.

To store the required quantities ($\tau^{max}$, $y_1^{max}$, $\tau^{min}$ and $y_1^{min}$), four accessory variables must be allocated to each thread. Initializing them with the initial condition via the corresponding pre-declared device function shown by the first code snippet in Sec.\,\ref{details_initialization_finalization} should cause no any problem. During the integration process, only the values of $\tau^{min}$ and $y_1^{min}$ need to be updated. The related code snippet is shown in the first listing in Sec.\,\ref{details_accessories}. The proper values of $\tau^{max}$ and $y_1^{max}$ are immediately set-up in the initialization procedure, see again Fig.\,\ref{BubbleCollapse}.

To be able to stop the integration at the next local maximum (end of the current iteration, initial state of the next one), a proper set-up of an event handling is required with the event function $F_{E1}=y_2$. The task is exactly the same as in case of the Duffing oscillator in the previous subsection. The tolerance is set again to $10^{-6}$. The event direction is $-1$ to detect only the local maxima, while the stop condition is $1$ to stop at the first local maximum. Keep in mind that if the initial condition is immediately inside an event zone (at a local maximum here) the corresponding event is not detected as it is discussed in Sec.\,\ref{concept_event_handling}.

The final set-up is the finalization process after every iterations. Since the forcing is quasiperiodic, in general, periodicity of the state space cannot be defined. Therefore, track of the time instances at the end of the integrations (let us denote by $t_E$ indicating the stop by event) is mandatory to be able to initialize the next iteration properly. More precisely, $t_0^{i+1}=t_E^i$, where $i$ is the serial number of the iterations. The suitable code snipped was already shown in Sec.\,\ref{details_initialization_finalization}. The final time instance of the overall simulation is not known in advance as the iterations are always stopped by the event. Thus, the end of the simulation time domain is set to a very high value $t_1=10^6$ to avoid incidental stops.

The present scan involves four parameters of the dual-frequency excitation; namely, the pressure amplitudes $P_{A1}$ and $P_{A2}$, and the frequencies $\omega_1$ and $\omega_2$. Their minimal and maximal values, their resolutions (how many values are taken) and the type of their distribution (linear or logarithmic) are summarized in Tab.\,\ref{tab:keller_miksis_parameters}. Observe that the number of the investigated parameters of each pressure amplitude is only two (only the minimum and the maximum). The overall number of scanned parameters is $2 \times 2 \times 128 \times 128 = 65536$. In each simulation, the first $1024$ iteration (collapses) are regarded as initial transients and discarded. The properties ($\tau^{max}$, $y_1^{max}$, $\tau^{min}$ and $y_1^{min}$) of the next $64$ collapses are registered and stored to datafile.

\begin{table}
\caption{Values of the control parameters of the four dimensional scan.}
\label{tab:keller_miksis_parameters}
\begin{tabular}{c|cccc}
\hline\noalign{\smallskip}
 & $P_{A1}$ & $P_{A2}$ & $\omega_1$ & $\omega_2$ \\
 & (bar)    & (bar)    & (kHz)      & (kHz)      \\
\noalign{\smallskip}\hline\noalign{\smallskip}
min.  & 0.5 & 0.7 & 20   & 20   \\
max.  & 1.1 & 1.2 & 1000 & 1000 \\
res.  & 2   & 2   & 128  & 128  \\
scale & lin & lin & log  & log  \\
\noalign{\smallskip}\hline
\end{tabular}
\end{table}

\begin{figure}[ht!]   
	\centering
		\subfloat{\includegraphics[width=8.4cm]{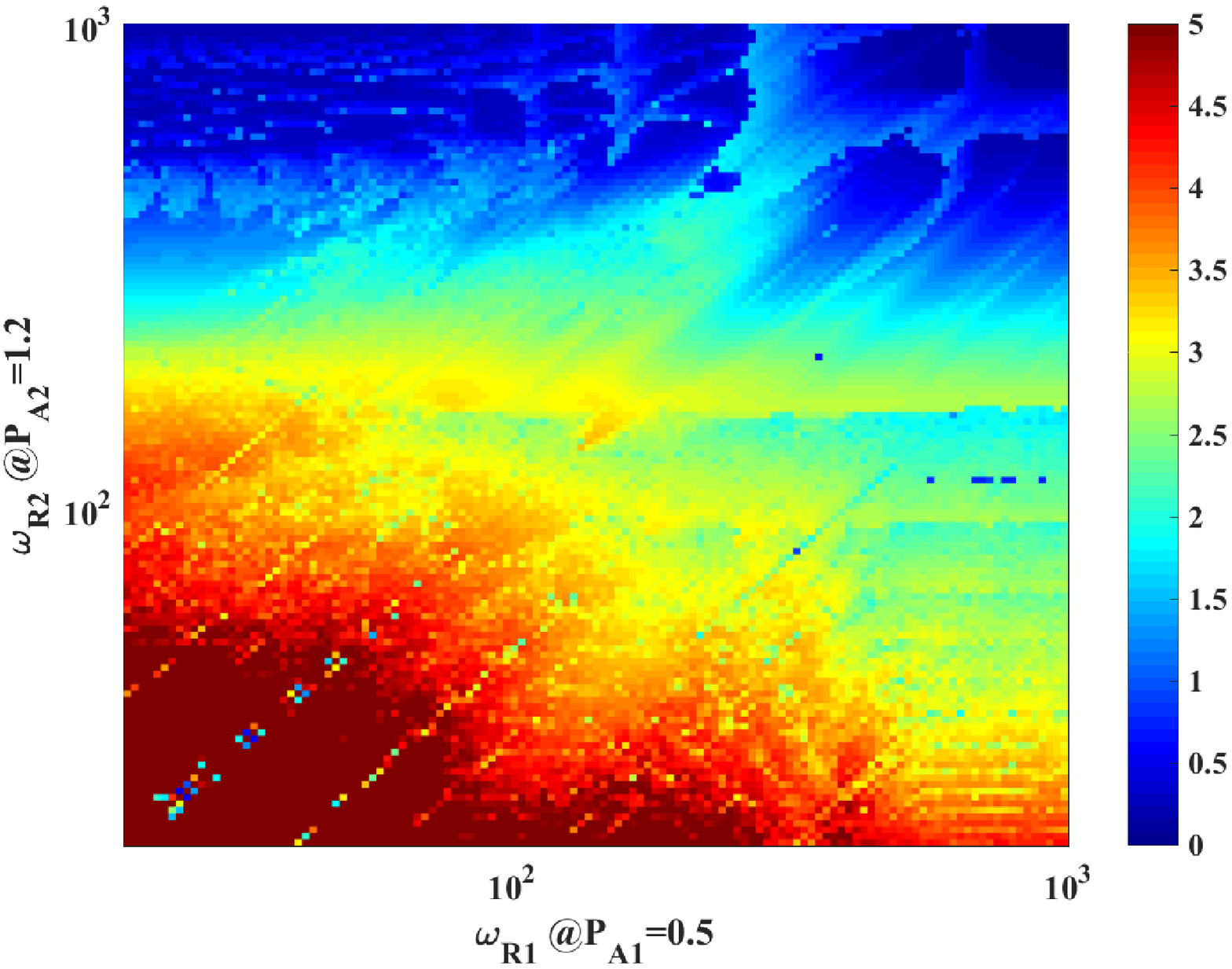}}
        \subfloat{\includegraphics[width=8.4cm]{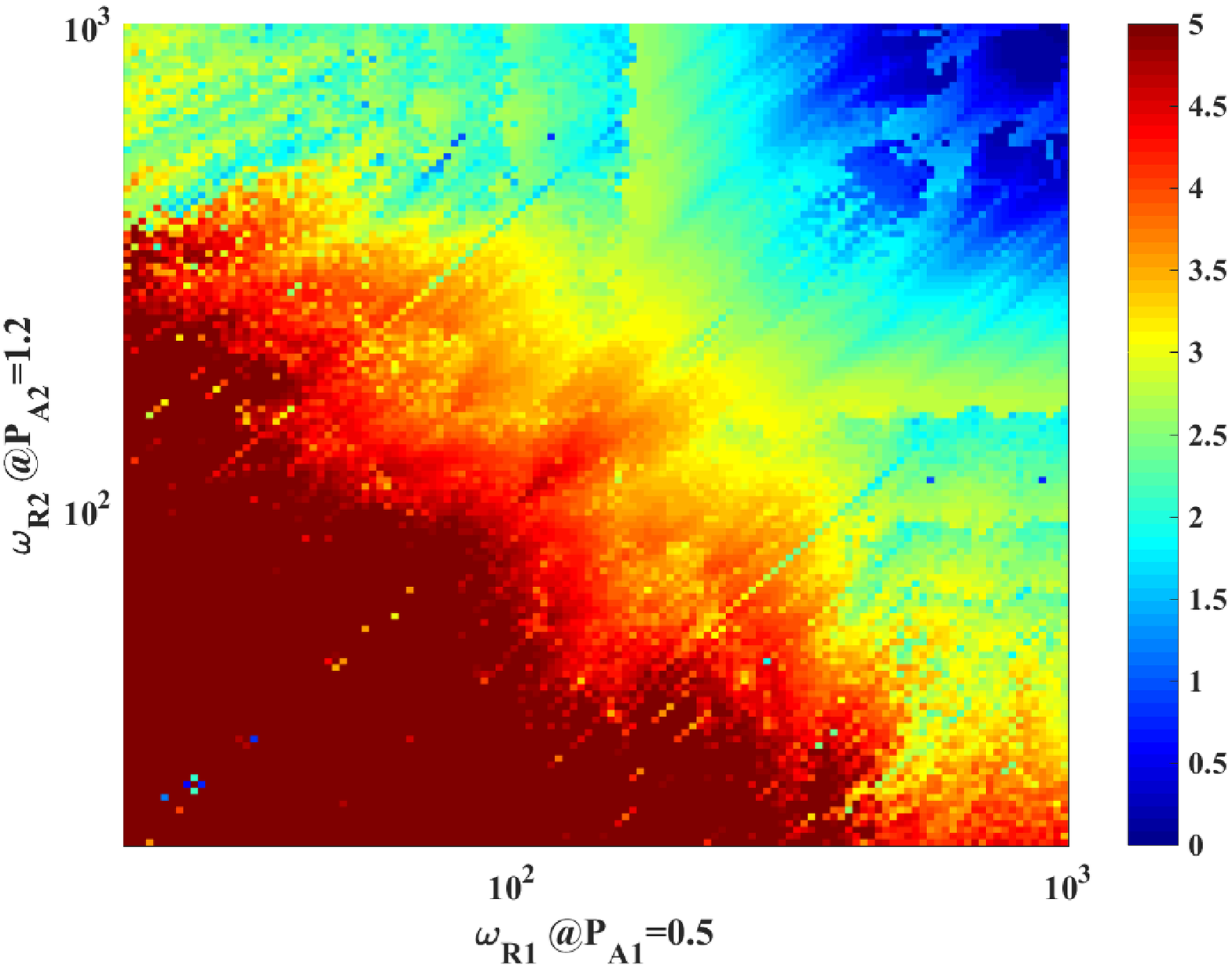}} \\
        \subfloat{\includegraphics[width=8.4cm]{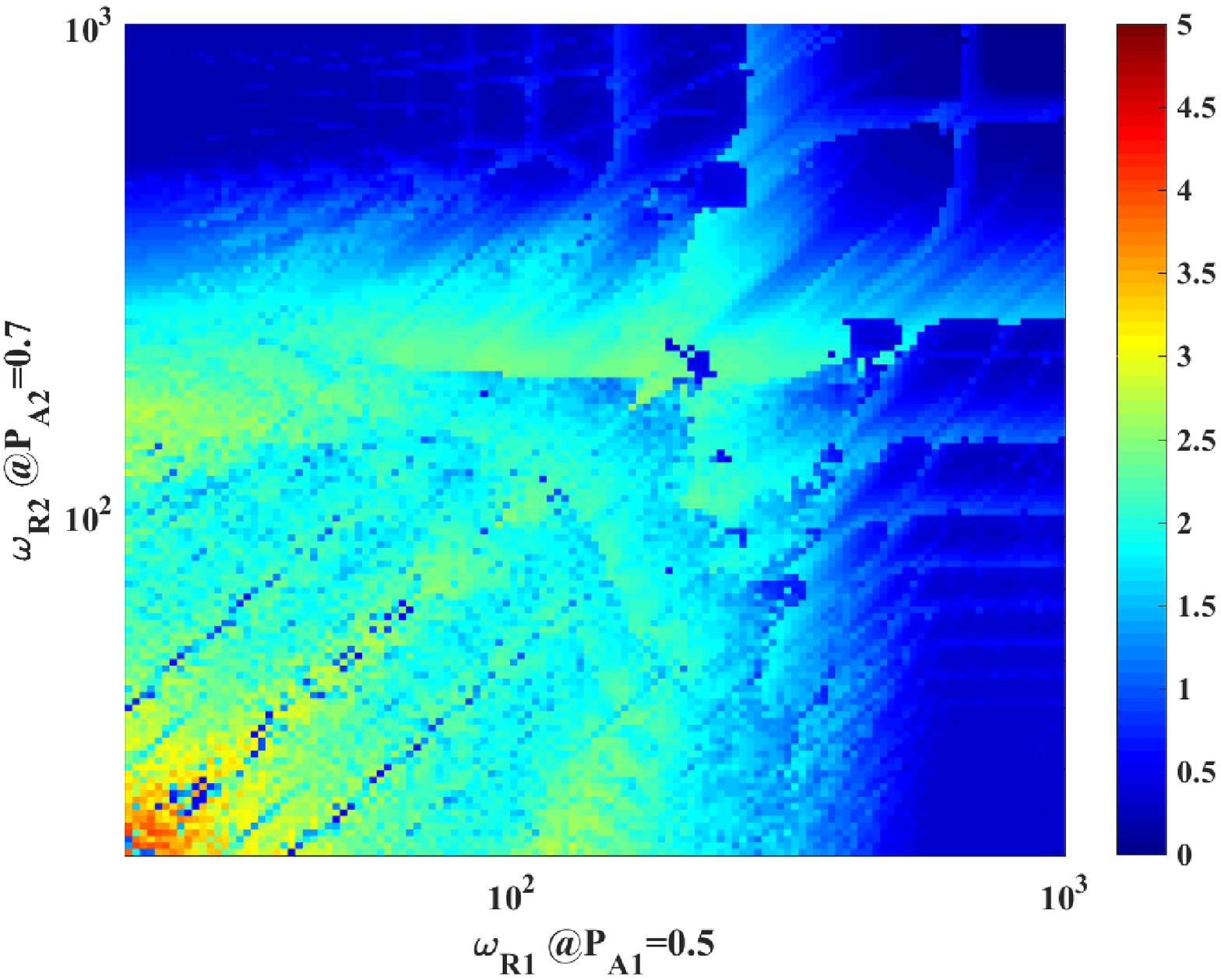}}
        \subfloat{\includegraphics[width=8.4cm]{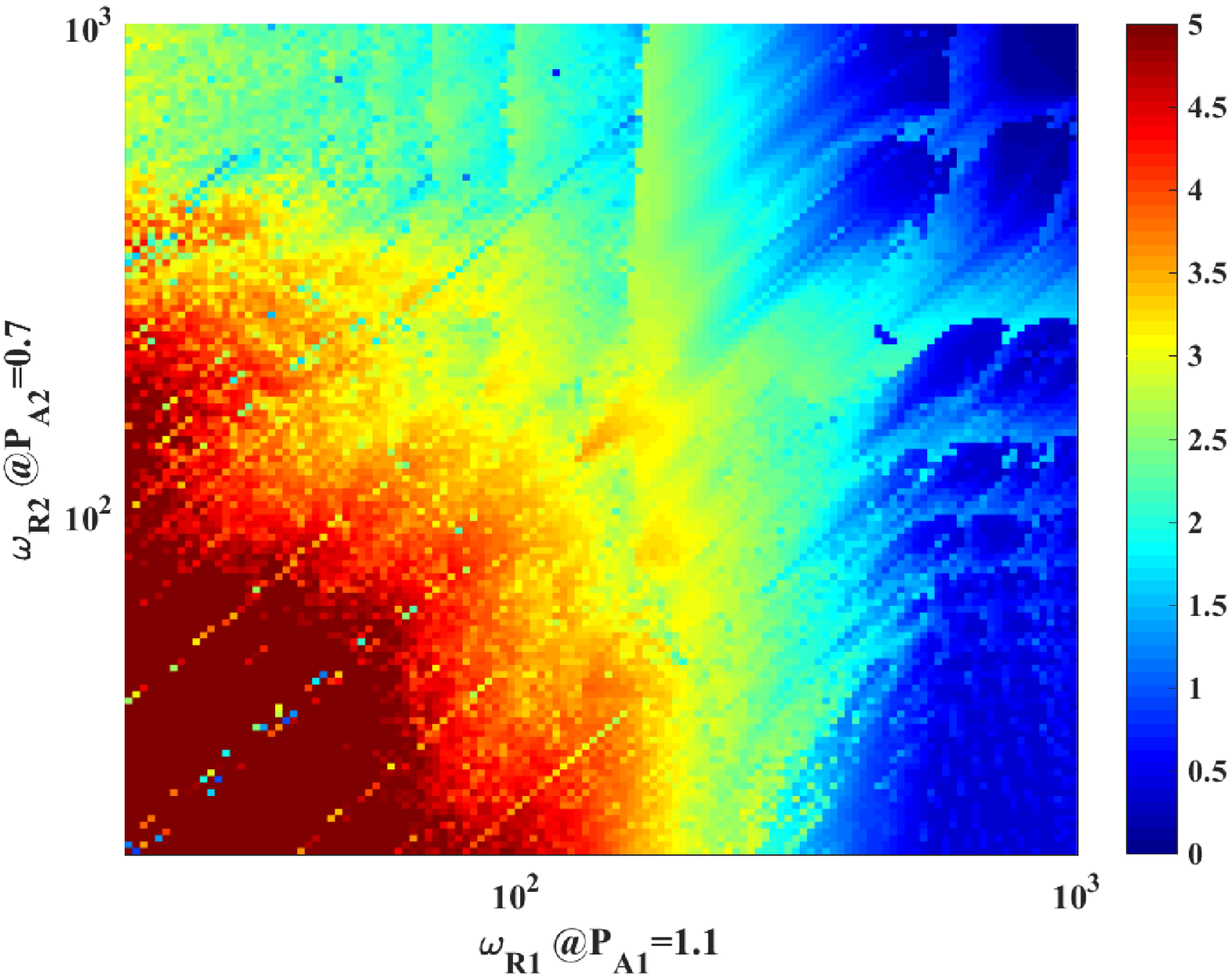}} \\
	\caption{Four dimensional parameter scan of the relative expansion of an oscillating single spherical bubble.}
	\label{Keller_Miksis_Collapse}
\end{figure}

In Fig.\,\ref{Keller_Miksis_Collapse}, the relative expansion ratio $y_{exp}$ is presented via four bi-parametric contour plots. The colour code indicates the magnitude of the relative expansion saturated at $y_{exp}=5$. The higher the value of $y_{exp}$ the stronger the bubble collapse (red domains). The control parameters in each subplot are the excitation frequencies $\omega_1$ and $\omega_2$. The pressure amplitude combinations are highlighted in the labels of the axes. At a given parameter set, the largest value out of the $64$ stored relative expansions is depicted.

\begin{table}
\caption{Summary of the profiling results of the bubble collapse test case corresponding to the kernel function of the Runge--Kutta--Cash--Karp solver. Both the absolute and relative tolerances of the adaptive time stepping are $10^{-10}$.}
\label{tab:nvvp_keller_miksis_RKCK45}
\begin{tabular}{ccccccccccc}
\hline\noalign{\smallskip}
$N_{reg}$ & $N_{t/b}$ & $N_T$ & $O_{TH}$ & $O_{SM}$ & $N_{b/SM}$ & $t_{c/t}$ & EW & AFU & FE   & MA  \\
          &           &       & (\%)     & (\%)     &            & ($ms$) &    &     & (\%) & (\%)\\
\noalign{\smallskip}\hline\noalign{\smallskip}
64 & 64 & 16384  & 50 & 37.9 & 17 & 96.8 & 4.77 & 8  & 31.2 & 89.4 \\
\noalign{\smallskip}\hline
\end{tabular}
\end{table}

The profiling is carried out only with a single kernel configuration presented in Tab.\,\ref{tab:nvvp_keller_miksis_RKCK45}. The simulations are performed for each subplot separately; therefore, the number of the threads during the profiling is actually $N_T=65536/4=16384$. The runtime of a kernel function is significantly higher compared to the previous test cases due to much more complex system and an order of magnitude stringent error tolerance. Nevertheless, the utilization of the arithmetic functions units $AFU$ and the multiprocessor activity $MA$ of the GPU is still relatively high. One reason for the lower values (compared to the Duffing oscillator) is the increased danger of thread divergence. The stronger the collapse the larger the difference in the time scales of a solution: compare the solutions near $y_1^{min}$ in Fig.\,\ref{BubbleCollapse} and near $y_1^{min3}$ in Fig.\,\ref{ConceptOfAccessories}. This means that trajectories exhibiting different collapse strength may require significantly different number of time steps which is the mayor source of thread divergence. One solution to ease this problem is to organize the problem so that the threads in a warp have similar parameter values. Thus, every thread in a warp may have similar collapse strength and similar amount of slow down during the collapse phase. Such a ``clustering'' technique is already suggested by \cite{Kroshko2013}. The number of registers required to avoid spilling in the Keller--Miksis test case is $184$.

\subsection{Test case: pressure relief valve (impact dynamics)} \label{test_impact_dynamics}

The final test case demonstrates the efficient handling of non-smooth dynamics through the simulation of a pressure relief valve introduced in Sec.\,\ref{model_pressure_relief_valve}. Since the system is autonomous, at lest two event functions have to be specified. One is for the definition of a suitable Poincar\'e section to be able to iterate the system from section to section. In this study, we choose $F_{E1}=y_2$ (local maximum of $y_1$) with stop condition $1$ (stop at every local maxima). Another event function is also needed to detect the impact between the valve body and the seat. It happens at $y_1=0$; thus, the second function is $F_{E2}=y_1$. In this case, the stop condition is $0$ (continue after impact) but the impact law must be incorporated into the corresponding pre-declared device function, see again Sec.\,\ref{details_event_handling}. Two accessories are used to store the maximum (Poincar\'e section) and the minimum (possible impact) of the valve position $y_1$ via the simple ``ordinary'' accessory device function. The main question in this section is that how efficient the utilization of the GPU is in case of the combination of multiple events and accessories. The assembled program code is a part of the supplementary material (PressureReliefValve.zip). Therefore, details as code snippets are omitted here. During the integration, the adaptive Runge--Kutta--Cash--Karp method are used with relative and absolute tolerances of $10^{-10}$.

\begin{figure}[ht!]   
	\centering
3		\includegraphics[width=8.6cm]{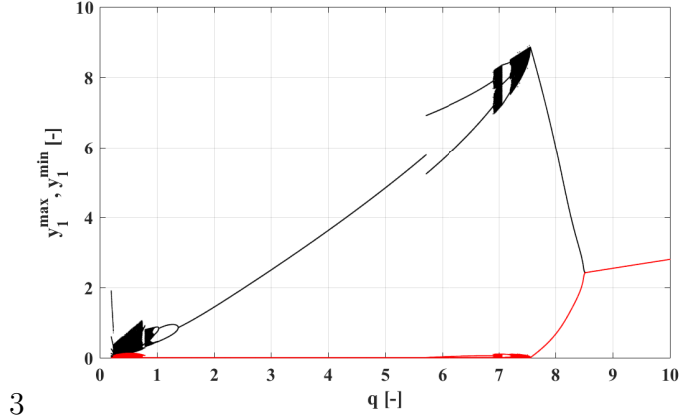}
	\caption{The maximum (black) and minimum (red) values of the valve position $y_1$ as a function of the dimensionless flow rate $q$ for the pressure relief valve model described in Sec.\,\ref{model_pressure_relief_valve}. The damping coefficient is $\kappa=1.25$, the precompression parameter is $\delta=10$ and the compressibility parameter is $\beta=20$. The number of the used threads and the resolution of the control parameter is $N_T=30720$.}
	\label{Pressure_Relief_Valve}
\end{figure}

The maxima $y_1^{max}$ (black dots) and minima $y_1^{min}$ (red dots) of the displacement of the valve body is depicted in Fig.\,\ref{Pressure_Relief_Valve} as a function of the dimensionless flow rate $q$ spanned between $0.2$ and $10$. The maxima are simply the Poincar\'e sections. The minima, however, are good indicator to show the range of parameters (approximately between $q=0.2$ and $q=7.5$) where impact dynamics occur ($y_1^{min}=0$). The figure shows good agreement with results of \cite{Hos2012}.

The performance characteristics of the kernel function are given in Tab.\,\ref{tab:nvvp_pressure_relief_valve_RKCK45} for two configurations. Although the multiprocessor activity $MA$ is still high, the arithmetic function utilization $AFU$ and the FLOP efficiency $FE$ is much lower then in the previous cases. This is due to the higher level of thread divergence. At high flow rates ($q>8.5$), where the red and black dots coincide, there is a single stable equilibrium the system can converge to. The trajectories converge to these points relatively fast, in which case the simulation stops very early (after $50$ time steps), see again Sec.\,\ref{concept_event_handling}. With decreasing flow rate, this equilibrium becomes unstable and a stable periodic orbit appear via a Hopf bifurcation. This is clearly indicated by the separation of the black and red dots. Near the hopf point, the amplitude of the orbits are small. Consequently, they exhibit no impact dynamics. However, as the flow rate decreases further, the amplitude and the period of the oscillation increases. Therefore, more time steps are necessary between two Poincar\'e sections in case of higher oscillation amplitudes. These two phenomena (equilibrium solution and increasing oscillation period) alone can be a high source of thread divergence. This is magnified when impact dynamics start to play a role in the system behaviour. In this case, the detection of an additional event is necessary. The number of registers need to avoid spilling in this final test case is $215$.

\begin{table}
\caption{Summary of the profiling results of the test case of the pressure relief valve corresponding to the kernel function of the Runge--Kutta--Cash--Karp solver. Both the absolute and relative tolerances of the adaptive time stepping are $10^{-10}$.}
\label{tab:nvvp_pressure_relief_valve_RKCK45}
\begin{tabular}{ccccccccccc}
\hline\noalign{\smallskip}
$N_{reg}$ & $N_{t/b}$ & $N_T$ & $O_{TH}$ & $O_{SM}$ & $N_{b/SM}$ & $t_{c/t}$ & EW & AFU & FE   & MA  \\
          &           &       & (\%)     & (\%)     &            & ($\mu s$) &    &     & (\%) & (\%)\\
\noalign{\smallskip}\hline\noalign{\smallskip}
64 & 64 & 15360  & 50 & 32.3 & 16 & 2.18 & 2.09 & 7  & 28.8 & 97.3 \\
64 & 64 & 30720  & 50 & 43.3 & 32 & 1.94 & 2.75 & 7  & 33.3 & 98.4 \\
\noalign{\smallskip}\hline
\end{tabular}
\end{table}

As a final remark, even if the utilization the GPU is somewhat lower than the previous test cases, the code can be still regarded as efficient. The unbalanced utilization is inherently due to the very heterogeneous feature of the solutions of the system. Such a behaviour would be difficult to handle on any other massively parallel platform. Nevertheless, the majority of the cheap processing power of the GPU is harnessed, which is a much better situation than using relatively expensive CPUs.


\section{Conclusion} \label{conclusion}

Throughout this paper, a general purpose and modular program package was introduced capable to solve huge number of independent ordinary differential equation systems in parallel. The framework of the code allows to handle multiple events efficiently and to store special features of the trajectories flexibly without the necessity to store every intermediate points of the solution. This makes the program fast by minimizing the pressure on global memory transactions. This efficiency was demonstrated via very different test cases in which the utilization of the processing power of the used GPU (NVIDIA Titan Black, Kepler architecture) was very high. For the profiling, the NVIDIA Visual Profiler (Release 7.5) was used. The available numerical schemes are the adaptive Runge--Kutta--Cash--Karp and the fifth order (fixed time step) Runge--Kutta schemes. In the forthcoming versions of the package, we intend to incorporate other explicit and even implicit numerical algorithms; and the efficient handling of global and diffusional couplings of many identical systems.


\section*{Acknowledgement}
The research reported in this paper was supported by the Higher Education Excellence Program of the Ministry of Human Capacities in the frame of Water science \& Disaster Prevention research area of Budapest University of Technology and Economics (BME FIKP-V\'IZ); and by the J\'anos Bolyai Research Scholarship of Hungarian Academy Sciences.


\section*{References}

\bibliographystyle{model1-num-names}
\biboptions{sort&compress}

\bibliography{2018_Hegedus_ComputPhysCommun.bib}

\end{document}